\documentclass[aps,prx,twocolumn,floatfix,longbibliography,superscriptaddress]{revtex4-2}

\usepackage{amsmath}
\usepackage{amssymb}
\usepackage{times}
\usepackage{braket}
\usepackage[pdftex]{graphicx}
\usepackage{color}
\usepackage{blindtext}
\usepackage{nicefrac}
\usepackage[colorlinks=true, citecolor=blue, urlcolor=blue, linkcolor=blue]{hyperref}

\usepackage{todonotes}

\renewcommand{\vec}[1]{\boldsymbol{#1}}

\renewcommand{\ket}[1]{\lvert#1\rangle} 
\newcommand{\braopket}[3]{\langle #1 | #2 | #3\rangle} 
\newcommand{\cone}{\mathrm{i}}

\begin{document}

\title{Orbital Hall effect and orbital edge states caused by $s$ electrons}

\author{Oliver Busch}
\email[Correspondence email address: ]{oliver.busch@physik.uni-halle.de}
\affiliation{Institut f\"ur Physik, Martin-Luther-Universit\"at Halle-Wittenberg, D-06099 Halle (Saale), Germany}

\author{Ingrid Mertig}
\affiliation{Institut f\"ur Physik, Martin-Luther-Universit\"at Halle-Wittenberg, D-06099 Halle (Saale), Germany}

\author{B{\"o}rge G{\"o}bel}
\email[Correspondence email address: ]{boerge.goebel@physik.uni-halle.de}
\affiliation{Institut f\"ur Physik, Martin-Luther-Universit\"at Halle-Wittenberg, D-06099 Halle (Saale), Germany}

\date{\today}

\begin{abstract}
An orbital current can be generated whenever an object has a translational and rotational degree of freedom. In condensed matter physics, intra-atomic contributions to the transverse orbital transport, labeled orbital Hall effect, rely on propagating wave packets that must consist of hybridized atomic orbitals. However, inter-atomic contributions have to be considered as well because they give rise to a new mechanism for generating orbital currents. As we show, even wave packets consisting purely of $s$ electrons can transport orbital angular momentum if they move on a cycloid trajectory. We introduce the kagome lattice with a single $s$ orbital per atom as the minimal model for the orbital Hall effect and observe the cycloid motion of the electrons in the surface states.
\end{abstract}

\maketitle

\section{Introduction}
The field of orbitronics is concerned with the orbital degree of freedom of electrons instead of their spin and charge~\cite{go2021orbitronics}. Despite the fact that orbital quenching~\cite{kittel2004surface} leads to a suppressed orbital magnetization in most solids, orbital currents often surpass spin currents in magnitude, as the latter require considerable spin-orbit coupling to be generated. This makes orbital currents highly attractive for dissipation-less orbitronic applications~\cite{cao2020prospect}.

Charge, spin and orbital currents can be generated by the charge~\cite{nagaosa2010anomalous}, spin~\cite{dyakonov1971current, hirsch1999spin, kato2004observation, sinova2015spin} and orbital Hall effects~\cite{zhang2005intrinsic, bernevig2005orbitronics, kontani2008giant, tanaka2008intrinsic, kontani2009giant}: The application of an electric field leads to the generation of the different types of currents as a transverse response. While the conventional (charge) Hall effect requires a broken inversion and time-reversal symmetry, the spin and orbital Hall effects can exist even in non-magnetic and centrosymmetric solids. The orbital Hall effect (OHE) has been predicted to exist even in systems without spin-orbit coupling but a hybridization of different atomic orbitals has been strictly required in the models up to now~\cite{go2018intrinsic, go2020orbital}.

The need for mixing of orbitals stems from the fact that earlier studies on the OHE were based on the atomic center approximation (ACA) for calculating the orbital angular momentum (OAM)~\cite{tanaka2008intrinsic, kontani2008giant, kontani2009giant, go2018intrinsic}: An OAM, that is supposed to be transported as an orbital current, can only be generated at a particular lattice site. However, since the building blocks of every solid are cubic harmonic orbitals ($s, p_x, p_y, p_z, d_{xy}, d_{yz}, d_{zx}, \ldots$), the OAM of a pure Bloch state always vanishes. The cubic harmonic orbitals need to hybridize (e.\,g. form the superpositions $p_x \pm \cone\, p_y$ or $d_{yz} \pm \cone\, d_{xz}$) in order to generate an OAM $L_z = m\hbar$. However, the ACA neglects inter-atomic contributions to the effect: A wave packet propagating across several lattice sites can carry an OAM irrespective of its orbital composition~\cite{chang1996berry}.

\begin{figure}[t!]
    \centering
    \includegraphics[width=\columnwidth]{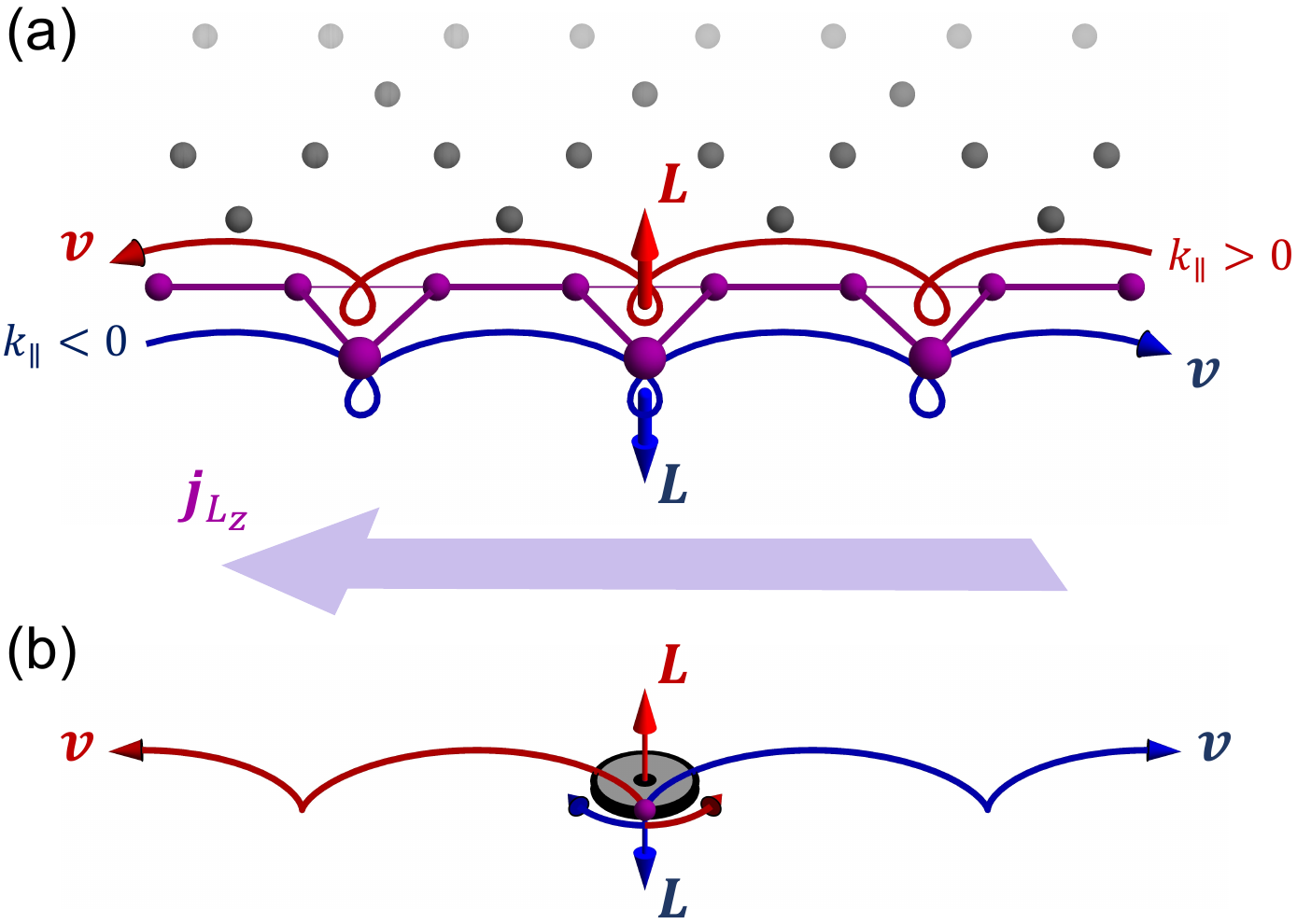}
    \caption{Schematic representation of orbital currents. (a) Orbital current  arising from edge states in a kagome lattice. Left- ($k_\parallel<0$) and right-propagating states ($k_\parallel>0$) carry an opposite orbital angular momentum ($L$) causing identical orbital currents $j_{L_z}$ and a compensated charge current. (b) Orbital current of a tire rolling forward ($v>0$) and backward ($v<0$). Both states carry an opposite orbital angular momentum causing identical orbital currents.}
    \label{fig:overview}
\end{figure}

In this paper, we take into account the inter-atomic contributions to the OHE by using the modern formulation of orbital magnetization~\cite{thonhauser2005orbital, xiao2005berry, shi2007quantum, yoda2018orbital, pezo2022orbital, cysne2022orbital, pezo2023orbital}. We show that the generation of an OHE does not require a specific orbital hybridization but can exist even for pure states. We propose a kagome lattice with only $s$ orbitals as the minimal model for the generation of an orbital Hall effect. We demonstrate that the OHE arises from a cycloid motion of a wave packet that is best observed in `geometrical'~\cite{yang2014emergent} edge states in a finite slab geometry. These states give rise to the same orbital current irrespective of their propagation direction [red and blue in Fig.~\ref{fig:overview}~(a)].

\section{Orbital currents in macroscopic systems} 
The ACA requires an often complicated hybridization of specific atomic orbitals for the generation of an OAM and an orbital magnetization~\cite{tanaka2008intrinsic, kontani2008giant, kontani2009giant, go2018intrinsic}. However, these quantities may emerge irrespective of the orbital contribution if the modern formulation of orbital magnetization is considered.

In the following, we want to point out that this is not surprising. Orbital currents are not `exotic' but appear whenever objects `translate' and `circulate' simultaneously. Even for macroscopic objects in classical mechanics, orbital currents are ubiquitous, as long as there is a translational and a rotational degree of freedom. For example, the rotors of a flying airplane can be identified with an orbital current. In the local coordinate system (moving with the center of mass) there is only a finite OAM $\vec{L}=\rho\int \vec{r}\times \vec{v}\,\mathrm{d}^3r$ since each point of the rotor follows a circular trajectory. In a global (stationary) coordinate system that lies on the path of the center of mass, the OAM is the same but it is transported by the translational motion of the airplane. The result is an orbital current.

As we will see later, most similar to the orbital current generated by the OHE in an $s$-orbital kagome model is the example of a rolling wheel (e.g. the tire of a car) [cf. Fig.~\ref{fig:overview}~(b)]. Without wheelslip, the friction force is strong enough to impose a constraint which couples the translational velocity (of the center of mass) $\vec{v}$ to the angular frequency $\vec{\omega}$ via the radius $R$: $v=R\omega$. Likewise, the angular momentum $\vec{L}\parallel \vec{\omega}$ is coupled to $\vec{v}$ via the moment of inertia $I$: $v=RL/I$. In particular, if $\vec{v}$ changes sign, $\vec{L}$ has to change as well. Since the orbital current $\vec{j}_L$ is proportional to $\vec{v}\times\vec{L}$, it is identical for forward and backward motion even though $\vec{L}$ reverses. 

Consequently, if one considers two wheels moving in opposite directions [red and blue in Fig.~\ref{fig:overview}~(b)], the total OAM $\vec{L}$ is compensated but a net orbital current $\vec{j}_\mathrm{L}$ arises. The trajectory of each point on the wheels is a cycloid in the global coordinate systems and a circle in the local coordinate system (origin is in the center of mass). As we demonstrate next, we observe the analogous scenario for $\vec{j}_\mathrm{L}$ and the same type of trajectory for a wave packet propagating at the edge of a kagome lattice~[cf.~Fig.~\ref{fig:overview}~(a)]. Here, the OAM $\vec{L}$ is coupled to the group velocity $\vec{v}=\nabla_{\vec{k}}\varepsilon/\hbar$.

\section{Orbital Hall effect} 
In the existing literature, the hybridization of different cubic orbitals was claimed to be the origin of the OHE~\cite{kontani2008giant, kontani2009giant, go2018intrinsic}. 
It was argued that they need to form the spherical harmonic orbitals with a net magnetic quantum number $m$ that gives rise to a finite OAM~\cite{go2018intrinsic, go2020orbital}. 
As discussed in the introduction, this is indeed the only possibility to generate an OHE when using the ACA. For example, a spherical atomic orbital with angular momentum quantum number $l=1$ and magnetic quantum number $m=+1$ is formed by the superposition of the cubic orbitals $p_x$ and $p_y$ as $\ket{l=1,m=1}=(\ket{p_x}-\cone\ket{p_y})/\sqrt{2}$ and gives rise to $L_z=m\hbar=+\hbar$. 

In order to be able to account for the transported OAM via a cycloid trajectory, it is not sufficient to use the ACA for calculating the OHE. Instead, we take into account intersite contributions via the modern formulation of orbital magnetization~\cite{thonhauser2005orbital, xiao2005berry, shi2007quantum, yoda2018orbital} and use it to calculate the OHE regardless of the orbital composition. The calculation is not based on the onsite OAM operator that accounts for the hybridization of the cubic atomic orbitals but it is calculated from the eigenvectors $\ket{\nu \vec{k}} \equiv \ket{\varphi_\nu(\vec{k})}$ and the eigenenergies $\varepsilon_{\nu \vec{k}} \equiv \varepsilon_{\nu}(\vec{k})$ of the tight-binding Hamiltonian that are $\vec{k}$ dependent.

The orbital Hall conductivity (OHC) $\sigma^{L_z}_{xy}$ quantifies the OHE by relating the generated orbital current to the applied electric field $j_{x}^{L_z}=\sigma^{L_z}_{xy}E_y$.
The OHC of a two-dimensional system at zero temperature located in the $xy$-plane, as considered in this work, can be computed as~\cite{kontani2009giant, go2018intrinsic, pezo2022orbital}
\begin{align}
    \sigma^{L_z}_{xy}(E_\text{F})= \frac{e}{\hbar}\sum_\nu \frac{1}{(2\pi)^2}\int_{\varepsilon_{\nu \vec{k}}\leq E_\text{F}}\Omega_{\nu,xy}^{L_z}(\vec{k}) \,\mathrm{d}^2k.\label{EQ:sigma_Lz_xy_Kubo}
\end{align}
for a specific Fermi level $E_\text{F}$. By analogy with the spin Hall conductivity, the OHC is given as the Brillouin zone integral of a `mixed Berry curvature' that is labeled `orbital Berry curvature'~\cite{pezo2022orbital}
\begin{align}
    \Omega_{\nu,xy}^{L_z}(\vec{k})= -2 \hbar^2\ \text{Im}\ \sum_{\mu\neq \nu} \frac{\braopket{\nu \vec{k}}{\Lambda_x^z}{\mu \vec{k}} \braopket{\mu \vec{k}}{v_y}{\nu \vec{k}}}{(\varepsilon_{\nu \vec{k}} - \varepsilon_{\mu \vec{k}})^2}.
\end{align}
 Herein, $v_i=\frac{1}{\hbar}\frac{\partial H}{\partial k_i}$ is the velocity operator that is calculated from the Hamiltonian $H$. The orbital current operator is given by $\Lambda_x^z \equiv \frac{1}{2}\{v_x ,L^z\}$ and can be evaluated according to the modern formulation of orbital magnetization via 
\begin{align*}
    \braopket{\nu \vec{k}}{\Lambda_x^z}{\mu \vec{k}} = \frac{1}{2}\sum_{\alpha} [ & \braopket{\nu \vec{k}}{v_x}{\alpha \vec{k}} \braopket{\alpha \vec{k}}{L_z}{\mu \vec{k}} \notag \\
    + & \braopket{\nu \vec{k}}{L_z}{\alpha \vec{k}} \braopket{\alpha \vec{k}}{v_x}{\mu \vec{k}} ].
\end{align*}
The matrix elements of the OAM comprise both diagonal elements and off-diagonal elements

\begin{align}
      & \braopket{\nu \vec{k}}{L_z}{\alpha \vec{k}} = -\cone  \frac{e\hbar^2}{4g_L\mu_\mathrm{B}}  \sum_{\beta \neq \nu, \alpha} \left( \frac{1}{\varepsilon_{\beta \vec{k}} - \varepsilon_{\nu \vec{k}}} + \frac{1}{\varepsilon_{\beta \vec{k}} - \varepsilon_{\alpha \vec{k}}} \right) \notag \\
     & \times \left(\braopket{\nu \vec{k}}{v_x}{\beta \vec{k}} \braopket{\beta \vec{k}}{v_y}{\alpha \vec{k}} - \braopket{\nu \vec{k}}{v_y}{\beta \vec{k}} \braopket{\beta \vec{k}}{v_x}{\alpha \vec{k}}\right)  \label{EQ:Lz_matrix_elements}
\end{align}

which is different compared to the ACA where $\braopket{\nu \vec{k}}{L_z}{\alpha \vec{k}}$ are constant matrix elements that mix different orbitals that are located at the same lattice site. The above equations were derived by Pezo \textit{et al.}~\cite{pezo2022orbital, pezo2023orbital} based on the modern formulation of the orbital magnetization in the language of wave packet dynamics~\cite{xiao2005berry}. The full derivation can be found in the main text and SM of Ref.~\cite{pezo2022orbital} but we have corrected a mistake in Eq.~\eqref{EQ:Lz_matrix_elements}~\footnote{We replaced `Im' (determining the imaginary part of the following expression) by the negative of the imaginary unit `$-\cone$'.}. Note that in a finite sample the derivation differs slightly: As presented in Ref.~\cite{go2017toward}, the velocity matrix elements can be calculated from the commutator between the position and the Hamilton operator.

\begin{figure*}[ht!]
    \centering
    \includegraphics[width=0.7\textwidth]{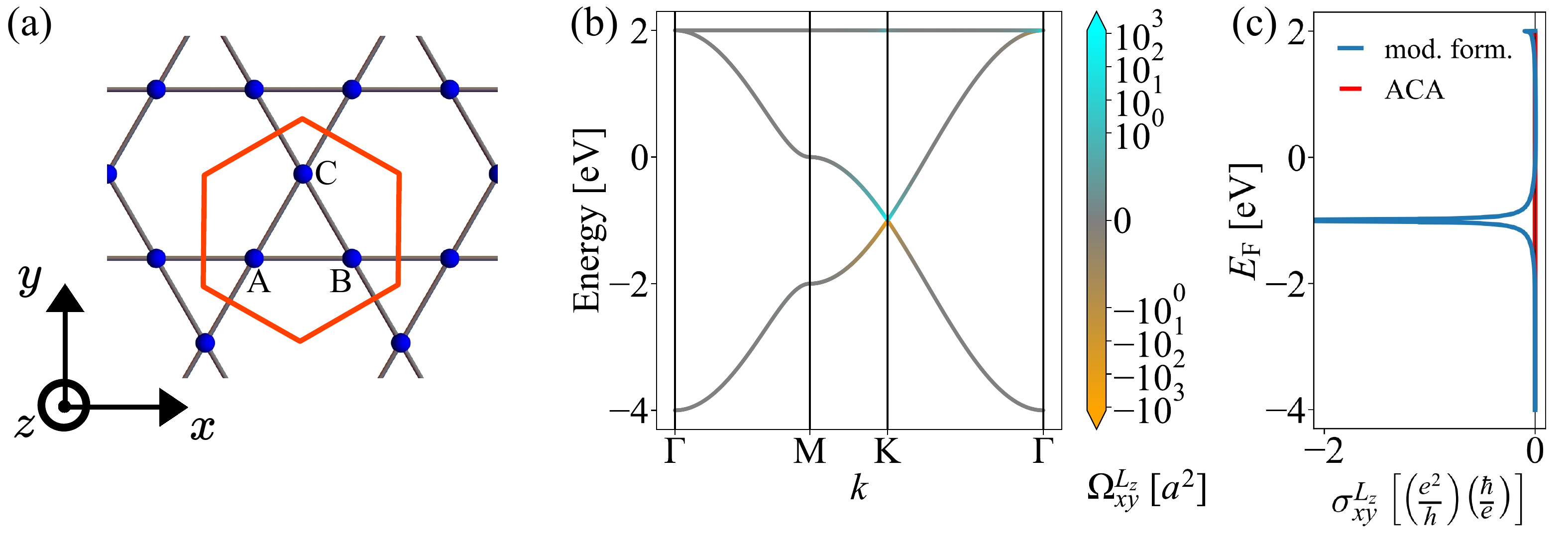}
    \caption{Orbital Hall effect in an $s$-orbital kagome lattice. (a) Kagome lattice with three basis atoms (A, B, C) per unit cell (red hexagon). (b) Band structure of the bulk system, where the color of the bands indicates the orbital Berry curvature $\Omega_{xy}^{L_z}$ using the modern formulation (cyan positive, orange negative). (c) Orbital Hall conductivity as a function of the Fermi level $E_\mathrm{F}$. Blue and red (zero) correspond to the modern formulation and ACA, respectively.}
    \label{fig:kagome_bulk}
\end{figure*}

\section{Minimal model: Kagome lattice with $s$ orbitals} 
As a minimal model that illustrates the crucial importance of using the modern formulations for computing the OHE, we have chosen a planar kagome lattice. This two-dimensional hexagonal lattice with a three-atomic unit cell [cf. Fig.~\ref{fig:kagome_bulk}~(a)] can be found in several materials including the ferromagnetic `kagome magnets' Fe$_3$Sn$_2$~\cite{Yin2018giant, Khadka2020anomalous} or Co$_3$Sn$_2$S$_2$~\cite{Liu2018giant, Yazyev2019an}, and the famous chiral antiferromagnets Mn$_3X$ ($X=$ Ir, Rh, Pt, Ge, Sn and Ga)~\cite{chen2014anomalous, nakatsuji2015large, nayak2016large, zhang2017strong}.

Since the main goal of our paper is to show that considering the ACA is not sufficient to calculate the OHE, we completely avoid any hybridization of orbitals by using one $s$ orbital per lattice site. Note that this orbital does not exhibit an onsite OAM (quantum number $l=0$ and so $m=0$). This means, only the modern formulation of the orbital magnetization contributes to the OHE via inter-atomic contributions. The ACA would always return a vanishing OHE which is why this formalism is inappropriate to quantify the effect.

Since $s$ orbitals do not exhibit spin-orbit coupling, our minimal model Hamiltonian includes only hopping terms $H= t\, \sum_{\braket{i, j}} c^\dagger_i\, c_j$ with the creation operator $c^\dagger_i$ and the annihilation operator $c_i$ of an electron at atom $i$. For simplicity, we consider only nearest-neighbor hopping and use $t=-1\,\mathrm{eV}$. 
In matrix form the Hamiltonian reads
\begin{align}
H=\begin{pmatrix}
0 & h_\mathrm{AB} & h_\mathrm{AC} \\
h_\mathrm{BA} & 0 & h_\mathrm{BC} \\
h_\mathrm{CA} & h_\mathrm{CB} & 0 \\
\end{pmatrix}
\end{align}
with $\vec{k}$-dependent entries $h_\mathrm{AB}=h_\mathrm{BA}=2t\, \cos\left(k_xa\right)$, $h_\mathrm{AC}=h_\mathrm{CA}=2t\, \cos\left(\frac{1}{2}k_xa+\frac{\sqrt{3}}{2}k_ya\right)$, and $h_\mathrm{BC}=h_\mathrm{CB}=2t\, \cos\left(-\frac{1}{2}k_xa+\frac{\sqrt{3}}{2}k_ya\right)$, and the lattice constant $a$. We consider spin so the matrix becomes $6\times 6$. The electronic band structure $\varepsilon_\nu(\vec{k})$ [cf. Fig.~\ref{fig:kagome_bulk}~(b)] exhibits Dirac points similar to the ones found in honeycomb lattices, like graphene~(cf.~Fig.~SM~1~(a),~(b) of the Supplemental Material~\cite{SupplMat}). Additionally, a flat band arises from the three-atomic basis that allows for closed loops and is not present in honeycomb lattices. All bands are spin degenerate due to inversion symmetry $\mathcal{I}$ and time-reversal symmetry $\mathcal{T}$ of the system in the absence of spin-orbit coupling.

As expected, the OHC vanishes in ACA because $s$ electrons do not carry an onsite OAM. However, in the modern formulation, the OHC is finite and depends on the location of the Fermi level [cf. Fig.~\ref{fig:kagome_bulk}~(c)]. 
The curve is almost constant within the bands but changes strongly at the band edges near the Dirac point and close to the flat band. For an occupation of 2 electrons per unit cell, the Fermi level is located at the Dirac point where the OHC is diverging.

Still, the effect cannot be attributed to the Dirac points alone, since a honeycomb lattice returns the same band structure (minus the flat band) but exhibits a vanishing OHC even in the modern formulation [cf. Fig.~SM~1~(c)~\cite{SupplMat}]. The reason why two almost identical band structures result in drastically different orbital Hall effects is that the OHC [Eq.~\eqref{EQ:sigma_Lz_xy_Kubo}] is determined by the eigenvectors as well. 

The onsite (diagonal) elements of the OAM $L_z$, calculated by Eq.~\eqref{EQ:Lz_matrix_elements}, vanish identically for all bands in the whole Brillouin zone (not shown here). However, the off-diagonal elements $\braopket{\nu \vec{k}}{L_z}{\alpha \vec{k}}$ are finite in the kagome system and enter the matrix elements of the orbital current operator $\Lambda_x^z$ and result in a finite orbital Berry curvature which is encoded as a color code (cyan positive, orange negative) in the electronic structure [cf.~Fig.~\ref{fig:kagome_bulk}~(b)]. This means no net OAM can be measured, even though the OHC is finite, and that the transported OAM is not generated at an individual lattice site but intersite contributions (off-diagonal elements) arise which correspond to a Bloch state that is spread out over all atoms in the unit cell. Thus, the situation is comparable to the spin Hall effect which can be finite despite a compensated spin magnetization $S$. 

Unfortunately, a trajectory of electron wave packets cannot be analyzed for the bulk system, as the tight-binding formalism only allows to access the probability density. 
However, investigating a slab geometry (nanoribbon) allows for a deeper insight into the origin of the orbital Hall effect. 

\section{Edge states generating cycloid trajectories}
A slab of the considered kagome system is only periodic along one direction giving rise to a single wave vector $k_\parallel$. Edges are introduced along the perpendicular direction. This gives rise to a large unit cell and many bands in the slab band structure. The vast majority of these bands represent a projection of the bulk band structure which is why signatures like the Dirac points and the flat band appear in the slab band structure as well. Note, that small super-cell gaps occur due to the finite width of the system and that they converge to zero in the limit of an infinitely wide nanoribbon. However, caused by the edges, we observe features that were absent in the bulk band structure; the so-called surface states.

\begin{figure}[ht!]
    \centering
    \includegraphics[width=\columnwidth]{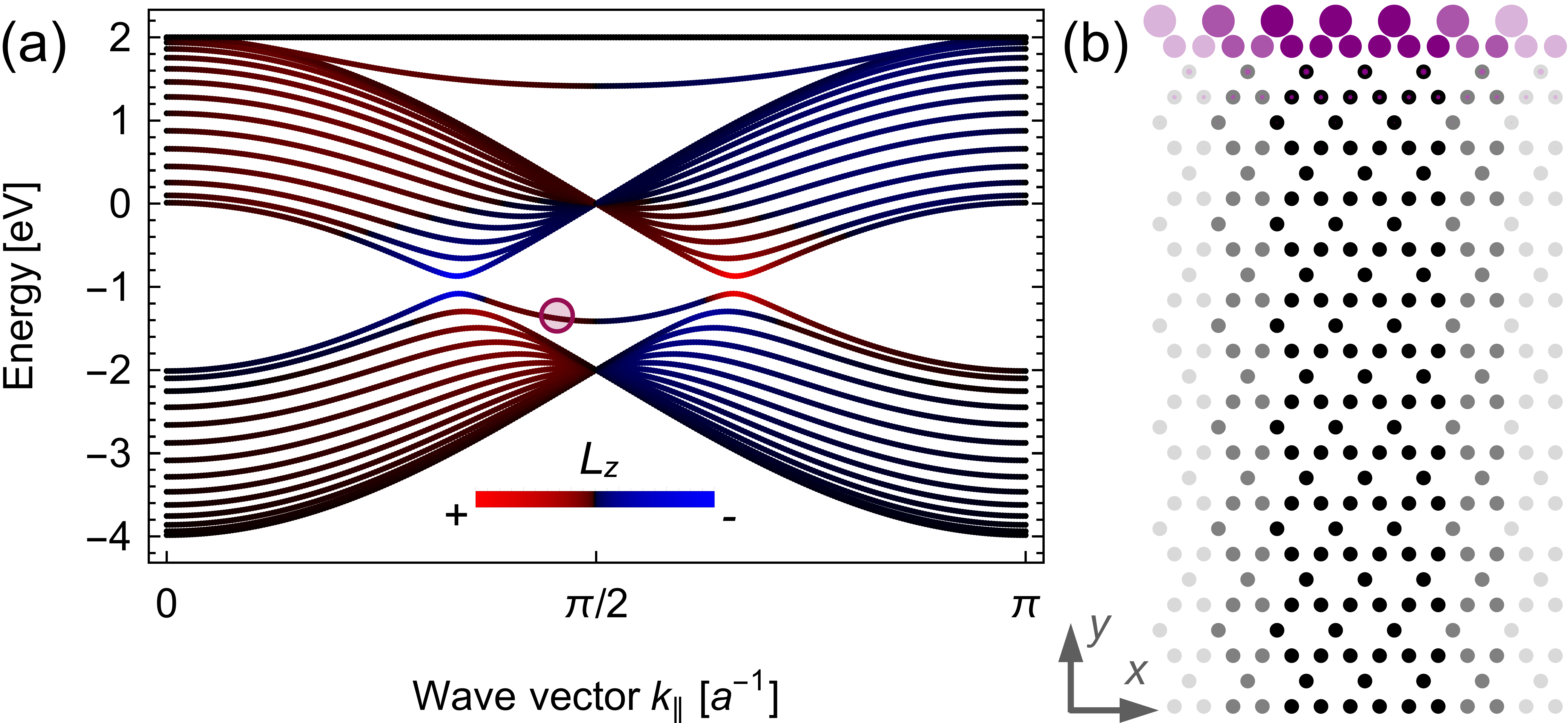}
    \caption{Edge states in an asymmetric kagome-lattice slab that is periodic along the $x$ direction. (a) Surface band structure along $k_{\parallel}$. 
    The color indicates positive (red) and negative (blue) values of $L_z$. (b) Considered slab with probability density (area of purple circles) of the surface state indicated in (a) for $k_{\parallel}=0.95\nicefrac{\pi}{2a}$.}
    \label{fig:kagome_edge_states}
\end{figure}

These states depend strongly on the specific shape of the edge: Most famously, a zigzag edge in a honeycomb lattice, like graphene, exhibits an edge state, while an armchair edge does not~\cite{niimi2005scanning, Kobayashi2005observation, Sasaki2006stabilization}. For the kagome lattice, we observe several edge states for multiple edge geometries and want to focus on the geometry presented in Fig.~\ref{fig:kagome_edge_states} which has one `straight' edge and one `triangular' edge [cf. bottom and top edges of the sketch in panel~(b)]. 

An edge state occurs close to the flat band and another between the two Dirac points [cf.~panel~(a)]. Their Chern numbers $C$ are zero due to the time-reversal symmetry of the system. In Ref.~\onlinecite{yang2014emergent} they have been labeled `geometrical edge states'. In contrast to the edge state in zigzag graphene, these states have a finite group velocity $v_{\nu}(k_\parallel)=\frac{1}{\hbar}\frac{\partial \varepsilon_{\nu}(k_\parallel)}{\partial k_\parallel}\neq 0$ allowing electron wave packets to propagate along the edges. 

The edge states are symmetric with respect to $k_\parallel$; for each right-propagating state at $+k_0$ there is a left-propagating state at $-k_0$. Due to the vanishing Chern numbers, the edges do not cause a quantum Hall effect. The $Z_2$ invariant and the spin Hall effect vanish as well because all the bands are spin degenerate due to the absence of spin-orbit coupling. 

However, the OAM $L_z$ is non-zero and has an opposite sign comparing states at $+k_0$ and $-k_0$. Note that $L_z$ has been zero in the bulk because due to the periodic repetition of the three-atomic unit cell, each upward-facing triangle automatically forms a neighboring downward-facing triangle with the same occupation. Therefore, each circular orbit in the bulk automatically generates an orbit with opposite circumferential direction resulting in a compensated $L_z$. At the triangular edge, this balance of upward- and downward-facing triangles is impaired and so a finite $L_z$ can be generated [color in Fig.~\ref{fig:kagome_edge_states}~(a)].

In Fig.~\ref{fig:kagome_edge_states}~(b) we show the probability density of electrons for the surface state between the Dirac points at $k_\parallel=0.95\frac{\pi}{2a}$.
Typical for an edge state, the probability density is largest for lattice sites close to the edge and decays exponentially going further into the bulk. In particular, the probability density is largest for the corners of the triangles at the very edge. Even though we can still only calculate the probability density and no currents, this distribution of electrons is in agreement with a cycloid trajectory as presented in Fig.~\ref{fig:overview}~(a). If a wave packet consists of $+k_0$ and $-k_0$ states, the scenario is in analogy with the forward and backward rolling wheel [cf. Fig.~\ref{fig:overview}~(b)], as discussed in the beginning of the paper: Edge states propagating along the right carry an opposite OAM compared to edge states propagating along the left, resulting in the same orbital current.

Our understanding of the behavior at the edge can be further condensed by considering a quasi-one-dimensional chain resembling the triangles at the very edges. Such a simplistic three-atomic model allows to calculate similar $k_\parallel$-resolved OAM and orbital currents as long as the chain does not have a glide-mirror symmetry with a symmetry axis along the periodic direction. In the Supplemental Material~\cite{SupplMat} we demonstrate that the asymmetric chain [Fig. SM~2~(b),~(c)~\cite{SupplMat}], as we find at the edge of the considered kagome lattice, leads to an orbital current. The cycloid trajectory with a curvature of constant sign can be understood as a superposition of a translation and a rotation as discussed before [cf.~Fig.~\ref{fig:overview}~(a)]. On the other hand, a symmetric zigzag chain cannot exhibit an orbital current since the probability density corresponds to a sine-like trajectory with alternating curvatures that compensate each other (Fig.~SM~2~(e),~(g)~\cite{SupplMat}) resulting in $\vec{L}=0$.
Note that in a honeycomb lattice, a symmetric zigzag-shaped edge leads to a dispersion-less surface state~\cite{Yao2009Edgestates} and therefore no orbital current can be generated. This also agrees with the finding that the OHE vanishes for a honeycomb lattice, like graphene, even in the modern theory. 

So far, we have only taken into account $s$ orbitals (which are equivalent to $p_z$ orbitals for two-dimensional systems located in the $xy$ plane). Taking into account all three $p$ orbitals causes hybridization of $p_x$ and $p_y$ which results in a finite OHE even within the ACA approach similar to the findings in Ref.~\onlinecite{go2018intrinsic}. OAM-polarized edge states arise in this case as well, as shown for a related material PtS$_2$ in Ref.~\onlinecite{costa2023connecting}. 
If spin-orbit coupling is considered, one observes a partial conversion of the OHE into the spin Hall effect as shown by Go \textit{et al.} using the ACA~\cite{go2018intrinsic}. Spin-orbit coupling lifts the spin degeneracy and the edge state splits up as shown by Sun \textit{et al.}~\cite{Sun2022spinvalley}. 

\begin{figure}[t!]
    \centering
    \includegraphics[width=\columnwidth]{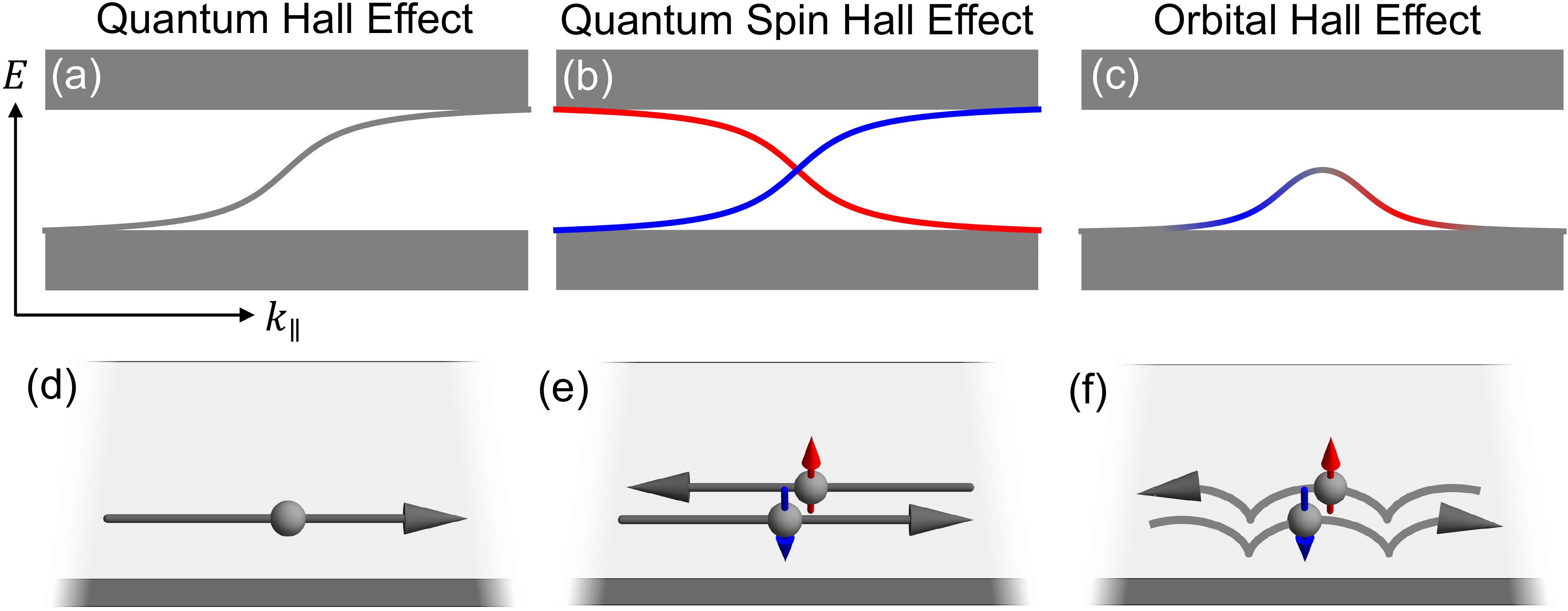}
    \caption{Comparison of edge states and corresponding edge transport. (a) Schematic band structure of a slab giving rise to the quantum Hall effect. The bulk conduction and valence bands (gray rectangles) are connected by a chiral edge state that is not spin polarized (gray). (b) Quantum spin Hall effect. The two edge states are spin polarized (red: up, blue: down). (c) Orbital Hall effect. The geometric edge state does not bridge the gap and is partially polarized with respect to the orbital angular momentum (red: positive, blue: negative). (d)-(f) Corresponding edge currents in real space. The colored arrows in (e) correspond to the spin and in (f) to the orbital angular momentum.}
    \label{fig:comparison}
\end{figure}

A categorization of the quantum Hall, quantum spin Hall and the edge signature of the orbital Hall effect can be found in Fig.~\ref{fig:comparison}. Even though their origin is fundamentally different, the spin Hall effect (SHE) and OHE exhibit similar transport of (spin or orbital) angular momentum not only in the bulk but also along the edge. The SHE caused by an edge state is quantized due to the complete spin polarization of the surface states. This is not the case for the OHE because $L$ is not quantized.
In contrast to the quantized versions of the charge and SHE, the orbital edge current 
is not protected. In the considered kagome system, the edge contribution to the OHE appears in its pure form and is carried by a geometrical edge state that does not bridge the gap between two bulk bands [cf. Fig.~\ref{fig:comparison}(c)]. However, especially when spin-orbit coupling is considered, the SHE and OHE may arise at the same time and the edge states can disperse differently. The gap may be bridged and the topological invariants $C$ and $Z_2$ may become non-zero integers giving rise to quantum Hall and quantum SHE that may be superimposed with the edge contribution to the OHE.

\section{Conclusion} 
In conclusion, we used the method derived in Ref.~\cite{pezo2022orbital} to account for inter-site contributions to the OHE. While it has been known that these contributions can drastically differ from the ACA,
our study demonstrates the emergence of a net OHE in a kagome lattice composed solely of $s$ orbitals in which the OHE based on the ACA is strictly zero. Importantly, we find that the OHE is a more prevalent phenomenon compared to the spin Hall effect and can appear without a specific orbital composition. The emergence of orbital currents is much easier and omnipresent than anticipated by the ACA. Furthermore, we identify edge contributions to the OHE. The existence of edge states in this lattice gives rise to wave packets following a cycloid trajectory, akin to the trajectory of a particle on a rolling wheel. These observations highlight the significance of the OHE in various materials and its relevance for dissipation-less orbitronic applications. 

The magnitude of the calculated effect is considerable. At 1/3 filling, the OHC is diverging in an ideal sample. In a realistic material, in which the Dirac point opens slightly, we expect values up to several $\frac{\mathrm{e}}{2\pi}$. Therefore, they are on the same order of magnitude as typical spin Hall conductivities and the orbital Hall conductivities that have been calculated before using the ACA for other materials such as $4d$ and $5d$-transition metals~\cite{tanaka2008intrinsic, kontani2009giant} and later $3d$ materials~\cite{jo2018gigantic} and Pt~\cite{go2018intrinsic}. Recently, the OHE has been observed experimentally in Ti using the magneto-optical Kerr effect in which intra-site contributions are already included~\cite{Choi2023observation}. Furthermore, orbital currents have recently been detected on ultrafast time scales by THz emission spectroscopy measurements~\cite{seifert2023time}.

Moreover, we note that our findings bear relevance to the valley Hall effect~\cite{bhowal2021orbital,cysne2021disentangling}, which has garnered significant attention in graphene research. In the context of the kagome lattice, the presence of distinct Dirac points at K and K', which differ in a physical quantity, gives rise to a Hall effect in that quantity. This unique characteristic is exemplified in our study, where the opposite orbital angular momentum associated with $k$ and $-k$ states [extrema of $L_z$ at the Dirac points in Fig.~\ref{fig:kagome_edge_states}~(a)] leads to the observed OHE.

Importantly, the OHE is not limited to a specific material. Rather, it is expected to manifest in kagome materials, as well as materials that allow for loops in the unit cell. This generality expands the potential avenues for exploring and harnessing the OHE in diverse systems. We have carried out additional calculations and find that it is present also in a square-octagon lattice (cf.~Fig.~SM~1~\cite{SupplMat}) which has four basis atoms forming a square in the unit cell.

\begin{acknowledgments}
This work is funded by the Deutsche Forschungsgemeinschaft (DFG, German Research Foundation) -- Project-ID 328545488 -- TRR~227, project~B04.
\end{acknowledgments}

O.B. and B.G. contributed equally to this work. O.B. performed the tight-binding and transport calculations for the bulk system; B.G. performed the edge state calculations. All authors discussed the results.

\bibliography{short,MyLibrary}

\begin{thebibliography}{46}%
\makeatletter
\providecommand \@ifxundefined [1]{%
 \@ifx{#1\undefined}
}%
\providecommand \@ifnum [1]{%
 \ifnum #1\expandafter \@firstoftwo
 \else \expandafter \@secondoftwo
 \fi
}%
\providecommand \@ifx [1]{%
 \ifx #1\expandafter \@firstoftwo
 \else \expandafter \@secondoftwo
 \fi
}%
\providecommand \natexlab [1]{#1}%
\providecommand \enquote  [1]{``#1''}%
\providecommand \bibnamefont  [1]{#1}%
\providecommand \bibfnamefont [1]{#1}%
\providecommand \citenamefont [1]{#1}%
\providecommand \href@noop [0]{\@secondoftwo}%
\providecommand \href [0]{\begingroup \@sanitize@url \@href}%
\providecommand \@href[1]{\@@startlink{#1}\@@href}%
\providecommand \@@href[1]{\endgroup#1\@@endlink}%
\providecommand \@sanitize@url [0]{\catcode `\\12\catcode `\$12\catcode
  `\&12\catcode `\#12\catcode `\^12\catcode `\_12\catcode `\%12\relax}%
\providecommand \@@startlink[1]{}%
\providecommand \@@endlink[0]{}%
\providecommand \url  [0]{\begingroup\@sanitize@url \@url }%
\providecommand \@url [1]{\endgroup\@href {#1}{\urlprefix }}%
\providecommand \urlprefix  [0]{URL }%
\providecommand \Eprint [0]{\href }%
\providecommand \doibase [0]{https://doi.org/}%
\providecommand \selectlanguage [0]{\@gobble}%
\providecommand \bibinfo  [0]{\@secondoftwo}%
\providecommand \bibfield  [0]{\@secondoftwo}%
\providecommand \translation [1]{[#1]}%
\providecommand \BibitemOpen [0]{}%
\providecommand \bibitemStop [0]{}%
\providecommand \bibitemNoStop [0]{.\EOS\space}%
\providecommand \EOS [0]{\spacefactor3000\relax}%
\providecommand \BibitemShut  [1]{\csname bibitem#1\endcsname}%
\let\auto@bib@innerbib\@empty
\bibitem [{\citenamefont {Go}\ \emph {et~al.}(2021)\citenamefont {Go},
  \citenamefont {Jo}, \citenamefont {Lee}, \citenamefont {Kl{\"a}ui},\ and\
  \citenamefont {Mokrousov}}]{go2021orbitronics}%
  \BibitemOpen
  \bibfield  {author} {\bibinfo {author} {\bibfnamefont {D.}~\bibnamefont
  {Go}}, \bibinfo {author} {\bibfnamefont {D.}~\bibnamefont {Jo}}, \bibinfo
  {author} {\bibfnamefont {H.-W.}\ \bibnamefont {Lee}}, \bibinfo {author}
  {\bibfnamefont {M.}~\bibnamefont {Kl{\"a}ui}},\ and\ \bibinfo {author}
  {\bibfnamefont {Y.}~\bibnamefont {Mokrousov}},\ }\bibfield  {title} {\bibinfo
  {title} {Orbitronics: Orbital currents in solids},\ }\href
  {https://doi.org/10.1209/0295-5075/ac2653} {\bibfield  {journal} {\bibinfo
  {journal} {Europhysics Letters}\ }\textbf {\bibinfo {volume} {135}},\
  \bibinfo {pages} {37001} (\bibinfo {year} {2021})}\BibitemShut {NoStop}%
\bibitem [{\citenamefont {Kittel}(2004)}]{kittel2004surface}%
  \BibitemOpen
  \bibfield  {author} {\bibinfo {author} {\bibfnamefont {C.}~\bibnamefont
  {Kittel}},\ }\bibfield  {title} {\bibinfo {title} {Surface and interface
  physics},\ }\href@noop {} {\bibfield  {journal} {\bibinfo  {journal}
  {Introduction to Solid State Physics; Wiley: New York, NY, USA}\ ,\ \bibinfo
  {pages} {487}} (\bibinfo {year} {2004})}\BibitemShut {NoStop}%
\bibitem [{\citenamefont {Cao}\ \emph {et~al.}(2020)\citenamefont {Cao},
  \citenamefont {Xing}, \citenamefont {Lin}, \citenamefont {Zhang},
  \citenamefont {Zheng},\ and\ \citenamefont {Wang}}]{cao2020prospect}%
  \BibitemOpen
  \bibfield  {author} {\bibinfo {author} {\bibfnamefont {Y.}~\bibnamefont
  {Cao}}, \bibinfo {author} {\bibfnamefont {G.}~\bibnamefont {Xing}}, \bibinfo
  {author} {\bibfnamefont {H.}~\bibnamefont {Lin}}, \bibinfo {author}
  {\bibfnamefont {N.}~\bibnamefont {Zhang}}, \bibinfo {author} {\bibfnamefont
  {H.}~\bibnamefont {Zheng}},\ and\ \bibinfo {author} {\bibfnamefont
  {K.}~\bibnamefont {Wang}},\ }\bibfield  {title} {\bibinfo {title} {Prospect
  of spin-orbitronic devices and their applications},\ }\href
  {https://doi.org/https://doi.org/10.1016/j.isci.2020.101614} {\bibfield
  {journal} {\bibinfo  {journal} {iScience}\ }\textbf {\bibinfo {volume}
  {23}},\ \bibinfo {pages} {101614} (\bibinfo {year} {2020})}\BibitemShut
  {NoStop}%
\bibitem [{\citenamefont {Nagaosa}\ \emph {et~al.}(2010)\citenamefont
  {Nagaosa}, \citenamefont {Sinova}, \citenamefont {Onoda}, \citenamefont
  {MacDonald},\ and\ \citenamefont {Ong}}]{nagaosa2010anomalous}%
  \BibitemOpen
  \bibfield  {author} {\bibinfo {author} {\bibfnamefont {N.}~\bibnamefont
  {Nagaosa}}, \bibinfo {author} {\bibfnamefont {J.}~\bibnamefont {Sinova}},
  \bibinfo {author} {\bibfnamefont {S.}~\bibnamefont {Onoda}}, \bibinfo
  {author} {\bibfnamefont {A.~H.}\ \bibnamefont {MacDonald}},\ and\ \bibinfo
  {author} {\bibfnamefont {N.~P.}\ \bibnamefont {Ong}},\ }\bibfield  {title}
  {\bibinfo {title} {Anomalous {H}all effect},\ }\href
  {https://doi.org/10.1103/RevModPhys.82.1539} {\bibfield  {journal} {\bibinfo
  {journal} {Reviews of Modern Physics}\ }\textbf {\bibinfo {volume} {82}},\
  \bibinfo {pages} {1539} (\bibinfo {year} {2010})}\BibitemShut {NoStop}%
\bibitem [{\citenamefont {D'yakonov}\ and\ \citenamefont
  {Perel}(1971)}]{dyakonov1971current}%
  \BibitemOpen
  \bibfield  {author} {\bibinfo {author} {\bibfnamefont {M.~I.}\ \bibnamefont
  {D'yakonov}}\ and\ \bibinfo {author} {\bibfnamefont {V.~I.}\ \bibnamefont
  {Perel}},\ }\bibfield  {title} {\bibinfo {title} {Current-induced spin
  orientation of electrons in semiconductors},\ }\href
  {https://doi.org/https://doi.org/10.1016/0375-9601(71)90196-4} {\bibfield
  {journal} {\bibinfo  {journal} {Physics Letters A}\ }\textbf {\bibinfo
  {volume} {35}},\ \bibinfo {pages} {459} (\bibinfo {year} {1971})}\BibitemShut
  {NoStop}%
\bibitem [{\citenamefont {Hirsch}(1999)}]{hirsch1999spin}%
  \BibitemOpen
  \bibfield  {author} {\bibinfo {author} {\bibfnamefont {J.~E.}\ \bibnamefont
  {Hirsch}},\ }\bibfield  {title} {\bibinfo {title} {Spin {H}all effect},\
  }\href {https://doi.org/10.1103/PhysRevLett.83.1834} {\bibfield  {journal}
  {\bibinfo  {journal} {Physical Review Letters}\ }\textbf {\bibinfo {volume}
  {83}},\ \bibinfo {pages} {1834} (\bibinfo {year} {1999})}\BibitemShut
  {NoStop}%
\bibitem [{\citenamefont {Kato}\ \emph {et~al.}(2004)\citenamefont {Kato},
  \citenamefont {Myers}, \citenamefont {Gossard},\ and\ \citenamefont
  {Awschalom}}]{kato2004observation}%
  \BibitemOpen
  \bibfield  {author} {\bibinfo {author} {\bibfnamefont {Y.~K.}\ \bibnamefont
  {Kato}}, \bibinfo {author} {\bibfnamefont {R.~C.}\ \bibnamefont {Myers}},
  \bibinfo {author} {\bibfnamefont {A.~C.}\ \bibnamefont {Gossard}},\ and\
  \bibinfo {author} {\bibfnamefont {D.~D.}\ \bibnamefont {Awschalom}},\
  }\bibfield  {title} {\bibinfo {title} {Observation of the spin {H}all effect
  in semiconductors},\ }\href {https://doi.org/10.1126/science.1105514}
  {\bibfield  {journal} {\bibinfo  {journal} {Science}\ }\textbf {\bibinfo
  {volume} {306}},\ \bibinfo {pages} {1910} (\bibinfo {year}
  {2004})}\BibitemShut {NoStop}%
\bibitem [{\citenamefont {Sinova}\ \emph {et~al.}(2015)\citenamefont {Sinova},
  \citenamefont {Valenzuela}, \citenamefont {Wunderlich}, \citenamefont
  {Back},\ and\ \citenamefont {Jungwirth}}]{sinova2015spin}%
  \BibitemOpen
  \bibfield  {author} {\bibinfo {author} {\bibfnamefont {J.}~\bibnamefont
  {Sinova}}, \bibinfo {author} {\bibfnamefont {S.~O.}\ \bibnamefont
  {Valenzuela}}, \bibinfo {author} {\bibfnamefont {J.}~\bibnamefont
  {Wunderlich}}, \bibinfo {author} {\bibfnamefont {C.}~\bibnamefont {Back}},\
  and\ \bibinfo {author} {\bibfnamefont {T.}~\bibnamefont {Jungwirth}},\
  }\bibfield  {title} {\bibinfo {title} {Spin {H}all effects},\ }\href
  {https://doi.org/10.1103/RevModPhys.87.1213} {\bibfield  {journal} {\bibinfo
  {journal} {Reviews of Modern Physics}\ }\textbf {\bibinfo {volume} {87}},\
  \bibinfo {pages} {1213} (\bibinfo {year} {2015})}\BibitemShut {NoStop}%
\bibitem [{\citenamefont {Zhang}\ and\ \citenamefont
  {Yang}(2005)}]{zhang2005intrinsic}%
  \BibitemOpen
  \bibfield  {author} {\bibinfo {author} {\bibfnamefont {S.}~\bibnamefont
  {Zhang}}\ and\ \bibinfo {author} {\bibfnamefont {Z.}~\bibnamefont {Yang}},\
  }\bibfield  {title} {\bibinfo {title} {Intrinsic spin and orbital angular
  momentum {H}all effect},\ }\href
  {https://doi.org/10.1103/PhysRevLett.94.066602} {\bibfield  {journal}
  {\bibinfo  {journal} {Physical Review Letters}\ }\textbf {\bibinfo {volume}
  {94}},\ \bibinfo {pages} {066602} (\bibinfo {year} {2005})}\BibitemShut
  {NoStop}%
\bibitem [{\citenamefont {Bernevig}\ \emph {et~al.}(2005)\citenamefont
  {Bernevig}, \citenamefont {Hughes},\ and\ \citenamefont
  {Zhang}}]{bernevig2005orbitronics}%
  \BibitemOpen
  \bibfield  {author} {\bibinfo {author} {\bibfnamefont {B.~A.}\ \bibnamefont
  {Bernevig}}, \bibinfo {author} {\bibfnamefont {T.~L.}\ \bibnamefont
  {Hughes}},\ and\ \bibinfo {author} {\bibfnamefont {S.-C.}\ \bibnamefont
  {Zhang}},\ }\bibfield  {title} {\bibinfo {title} {Orbitronics: The intrinsic
  orbital current in $p$-doped silicon},\ }\href
  {https://doi.org/10.1103/PhysRevLett.95.066601} {\bibfield  {journal}
  {\bibinfo  {journal} {Physical Review Letters}\ }\textbf {\bibinfo {volume}
  {95}},\ \bibinfo {pages} {066601} (\bibinfo {year} {2005})}\BibitemShut
  {NoStop}%
\bibitem [{\citenamefont {Kontani}\ \emph {et~al.}(2008)\citenamefont
  {Kontani}, \citenamefont {Tanaka}, \citenamefont {Hirashima}, \citenamefont
  {Yamada},\ and\ \citenamefont {Inoue}}]{kontani2008giant}%
  \BibitemOpen
  \bibfield  {author} {\bibinfo {author} {\bibfnamefont {H.}~\bibnamefont
  {Kontani}}, \bibinfo {author} {\bibfnamefont {T.}~\bibnamefont {Tanaka}},
  \bibinfo {author} {\bibfnamefont {D.}~\bibnamefont {Hirashima}}, \bibinfo
  {author} {\bibfnamefont {K.}~\bibnamefont {Yamada}},\ and\ \bibinfo {author}
  {\bibfnamefont {J.}~\bibnamefont {Inoue}},\ }\bibfield  {title} {\bibinfo
  {title} {Giant intrinsic spin and orbital {H}all effects in {Sr$_2M$O$_4$
  ($M$= Ru, Rh, Mo)}},\ }\href {https://doi.org/10.1103/PhysRevLett.100.096601}
  {\bibfield  {journal} {\bibinfo  {journal} {Physical Review Letters}\
  }\textbf {\bibinfo {volume} {100}},\ \bibinfo {pages} {096601} (\bibinfo
  {year} {2008})}\BibitemShut {NoStop}%
\bibitem [{\citenamefont {Tanaka}\ \emph {et~al.}(2008)\citenamefont {Tanaka},
  \citenamefont {Kontani}, \citenamefont {Naito}, \citenamefont {Naito},
  \citenamefont {Hirashima}, \citenamefont {Yamada},\ and\ \citenamefont
  {Inoue}}]{tanaka2008intrinsic}%
  \BibitemOpen
  \bibfield  {author} {\bibinfo {author} {\bibfnamefont {T.}~\bibnamefont
  {Tanaka}}, \bibinfo {author} {\bibfnamefont {H.}~\bibnamefont {Kontani}},
  \bibinfo {author} {\bibfnamefont {M.}~\bibnamefont {Naito}}, \bibinfo
  {author} {\bibfnamefont {T.}~\bibnamefont {Naito}}, \bibinfo {author}
  {\bibfnamefont {D.~S.}\ \bibnamefont {Hirashima}}, \bibinfo {author}
  {\bibfnamefont {K.}~\bibnamefont {Yamada}},\ and\ \bibinfo {author}
  {\bibfnamefont {J.}~\bibnamefont {Inoue}},\ }\bibfield  {title} {\bibinfo
  {title} {Intrinsic spin {H}all effect and orbital {H}all effect in {$4d$ and
  $5d$} transition metals},\ }\href
  {https://doi.org/10.1103/PhysRevB.77.165117} {\bibfield  {journal} {\bibinfo
  {journal} {Physical Review B}\ }\textbf {\bibinfo {volume} {77}},\ \bibinfo
  {pages} {165117} (\bibinfo {year} {2008})}\BibitemShut {NoStop}%
\bibitem [{\citenamefont {Kontani}\ \emph {et~al.}(2009)\citenamefont
  {Kontani}, \citenamefont {Tanaka}, \citenamefont {Hirashima}, \citenamefont
  {Yamada},\ and\ \citenamefont {Inoue}}]{kontani2009giant}%
  \BibitemOpen
  \bibfield  {author} {\bibinfo {author} {\bibfnamefont {H.}~\bibnamefont
  {Kontani}}, \bibinfo {author} {\bibfnamefont {T.}~\bibnamefont {Tanaka}},
  \bibinfo {author} {\bibfnamefont {D.}~\bibnamefont {Hirashima}}, \bibinfo
  {author} {\bibfnamefont {K.}~\bibnamefont {Yamada}},\ and\ \bibinfo {author}
  {\bibfnamefont {J.}~\bibnamefont {Inoue}},\ }\bibfield  {title} {\bibinfo
  {title} {Giant orbital {H}all effect in transition metals: Origin of large
  spin and anomalous {H}all effects},\ }\href
  {https://doi.org/10.1103/PhysRevLett.102.016601} {\bibfield  {journal}
  {\bibinfo  {journal} {Physical Review Letters}\ }\textbf {\bibinfo {volume}
  {102}},\ \bibinfo {pages} {016601} (\bibinfo {year} {2009})}\BibitemShut
  {NoStop}%
\bibitem [{\citenamefont {Go}\ \emph {et~al.}(2018)\citenamefont {Go},
  \citenamefont {Jo}, \citenamefont {Kim},\ and\ \citenamefont
  {Lee}}]{go2018intrinsic}%
  \BibitemOpen
  \bibfield  {author} {\bibinfo {author} {\bibfnamefont {D.}~\bibnamefont
  {Go}}, \bibinfo {author} {\bibfnamefont {D.}~\bibnamefont {Jo}}, \bibinfo
  {author} {\bibfnamefont {C.}~\bibnamefont {Kim}},\ and\ \bibinfo {author}
  {\bibfnamefont {H.-W.}\ \bibnamefont {Lee}},\ }\bibfield  {title} {\bibinfo
  {title} {Intrinsic spin and orbital {H}all effects from orbital texture},\
  }\href {https://doi.org/10.1103/PhysRevLett.121.086602} {\bibfield  {journal}
  {\bibinfo  {journal} {Physical Review Letters}\ }\textbf {\bibinfo {volume}
  {121}},\ \bibinfo {pages} {086602} (\bibinfo {year} {2018})}\BibitemShut
  {NoStop}%
\bibitem [{\citenamefont {Go}\ and\ \citenamefont {Lee}(2020)}]{go2020orbital}%
  \BibitemOpen
  \bibfield  {author} {\bibinfo {author} {\bibfnamefont {D.}~\bibnamefont
  {Go}}\ and\ \bibinfo {author} {\bibfnamefont {H.-W.}\ \bibnamefont {Lee}},\
  }\bibfield  {title} {\bibinfo {title} {Orbital torque: Torque generation by
  orbital current injection},\ }\href
  {https://doi.org/10.1103/PhysRevResearch.2.013177} {\bibfield  {journal}
  {\bibinfo  {journal} {Physical Review Research}\ }\textbf {\bibinfo {volume}
  {2}},\ \bibinfo {pages} {013177} (\bibinfo {year} {2020})}\BibitemShut
  {NoStop}%
\bibitem [{\citenamefont {Chang}\ and\ \citenamefont
  {Niu}(1996)}]{chang1996berry}%
  \BibitemOpen
  \bibfield  {author} {\bibinfo {author} {\bibfnamefont {M.-C.}\ \bibnamefont
  {Chang}}\ and\ \bibinfo {author} {\bibfnamefont {Q.}~\bibnamefont {Niu}},\
  }\bibfield  {title} {\bibinfo {title} {Berry phase, hyperorbits, and the
  {H}ofstadter spectrum: Semiclassical dynamics in magnetic {B}loch bands},\
  }\href {https://doi.org/10.1103/PhysRevB.53.7010} {\bibfield  {journal}
  {\bibinfo  {journal} {Physical Review B}\ }\textbf {\bibinfo {volume} {53}},\
  \bibinfo {pages} {7010} (\bibinfo {year} {1996})}\BibitemShut {NoStop}%
\bibitem [{\citenamefont {Thonhauser}\ \emph {et~al.}(2005)\citenamefont
  {Thonhauser}, \citenamefont {Ceresoli}, \citenamefont {Vanderbilt},\ and\
  \citenamefont {Resta}}]{thonhauser2005orbital}%
  \BibitemOpen
  \bibfield  {author} {\bibinfo {author} {\bibfnamefont {T.}~\bibnamefont
  {Thonhauser}}, \bibinfo {author} {\bibfnamefont {D.}~\bibnamefont
  {Ceresoli}}, \bibinfo {author} {\bibfnamefont {D.}~\bibnamefont
  {Vanderbilt}},\ and\ \bibinfo {author} {\bibfnamefont {R.}~\bibnamefont
  {Resta}},\ }\bibfield  {title} {\bibinfo {title} {Orbital magnetization in
  periodic insulators},\ }\href {https://doi.org/10.1103/PhysRevLett.95.137205}
  {\bibfield  {journal} {\bibinfo  {journal} {Physical Review Letters}\
  }\textbf {\bibinfo {volume} {95}},\ \bibinfo {pages} {137205} (\bibinfo
  {year} {2005})}\BibitemShut {NoStop}%
\bibitem [{\citenamefont {Xiao}\ \emph {et~al.}(2005)\citenamefont {Xiao},
  \citenamefont {Shi},\ and\ \citenamefont {Niu}}]{xiao2005berry}%
  \BibitemOpen
  \bibfield  {author} {\bibinfo {author} {\bibfnamefont {D.}~\bibnamefont
  {Xiao}}, \bibinfo {author} {\bibfnamefont {J.}~\bibnamefont {Shi}},\ and\
  \bibinfo {author} {\bibfnamefont {Q.}~\bibnamefont {Niu}},\ }\bibfield
  {title} {\bibinfo {title} {Berry phase correction to electron density of
  states in solids},\ }\href {https://doi.org/10.1103/PhysRevLett.95.137204}
  {\bibfield  {journal} {\bibinfo  {journal} {Physical Review Letters}\
  }\textbf {\bibinfo {volume} {95}},\ \bibinfo {pages} {137204} (\bibinfo
  {year} {2005})}\BibitemShut {NoStop}%
\bibitem [{\citenamefont {Shi}\ \emph {et~al.}(2007)\citenamefont {Shi},
  \citenamefont {Vignale}, \citenamefont {Xiao},\ and\ \citenamefont
  {Niu}}]{shi2007quantum}%
  \BibitemOpen
  \bibfield  {author} {\bibinfo {author} {\bibfnamefont {J.}~\bibnamefont
  {Shi}}, \bibinfo {author} {\bibfnamefont {G.}~\bibnamefont {Vignale}},
  \bibinfo {author} {\bibfnamefont {D.}~\bibnamefont {Xiao}},\ and\ \bibinfo
  {author} {\bibfnamefont {Q.}~\bibnamefont {Niu}},\ }\bibfield  {title}
  {\bibinfo {title} {Quantum theory of orbital magnetization and its
  generalization to interacting systems},\ }\href
  {https://doi.org/10.1103/PhysRevLett.99.197202} {\bibfield  {journal}
  {\bibinfo  {journal} {Physical Review Letters}\ }\textbf {\bibinfo {volume}
  {99}},\ \bibinfo {pages} {197202} (\bibinfo {year} {2007})}\BibitemShut
  {NoStop}%
\bibitem [{\citenamefont {Yoda}\ \emph {et~al.}(2018)\citenamefont {Yoda},
  \citenamefont {Yokoyama},\ and\ \citenamefont {Murakami}}]{yoda2018orbital}%
  \BibitemOpen
  \bibfield  {author} {\bibinfo {author} {\bibfnamefont {T.}~\bibnamefont
  {Yoda}}, \bibinfo {author} {\bibfnamefont {T.}~\bibnamefont {Yokoyama}},\
  and\ \bibinfo {author} {\bibfnamefont {S.}~\bibnamefont {Murakami}},\
  }\bibfield  {title} {\bibinfo {title} {Orbital {E}delstein effect as a
  condensed-matter analog of solenoids},\ }\href
  {https://doi.org/10.1021/acs.nanolett.7b04300} {\bibfield  {journal}
  {\bibinfo  {journal} {Nano Letters}\ }\textbf {\bibinfo {volume} {18}},\
  \bibinfo {pages} {916} (\bibinfo {year} {2018})}\BibitemShut {NoStop}%
\bibitem [{\citenamefont {Pezo}\ \emph {et~al.}(2022)\citenamefont {Pezo},
  \citenamefont {Ovalle},\ and\ \citenamefont {Manchon}}]{pezo2022orbital}%
  \BibitemOpen
  \bibfield  {author} {\bibinfo {author} {\bibfnamefont {A.}~\bibnamefont
  {Pezo}}, \bibinfo {author} {\bibfnamefont {D.~G.}\ \bibnamefont {Ovalle}},\
  and\ \bibinfo {author} {\bibfnamefont {A.}~\bibnamefont {Manchon}},\
  }\bibfield  {title} {\bibinfo {title} {Orbital {H}all effect in crystals:
  {I}nteratomic versus intra-atomic contributions},\ }\href
  {https://doi.org/10.1103/PhysRevB.106.104414} {\bibfield  {journal} {\bibinfo
   {journal} {Physical Review B}\ }\textbf {\bibinfo {volume} {106}},\ \bibinfo
  {pages} {104414} (\bibinfo {year} {2022})}\BibitemShut {NoStop}%
\bibitem [{\citenamefont {Cysne}\ \emph {et~al.}(2022)\citenamefont {Cysne},
  \citenamefont {Bhowal}, \citenamefont {Vignale},\ and\ \citenamefont
  {Rappoport}}]{cysne2022orbital}%
  \BibitemOpen
  \bibfield  {author} {\bibinfo {author} {\bibfnamefont {T.~P.}\ \bibnamefont
  {Cysne}}, \bibinfo {author} {\bibfnamefont {S.}~\bibnamefont {Bhowal}},
  \bibinfo {author} {\bibfnamefont {G.}~\bibnamefont {Vignale}},\ and\ \bibinfo
  {author} {\bibfnamefont {T.~G.}\ \bibnamefont {Rappoport}},\ }\bibfield
  {title} {\bibinfo {title} {Orbital {H}all effect in bilayer transition metal
  dichalcogenides: From the intra-atomic approximation to the {B}loch states
  orbital magnetic moment approach},\ }\href
  {https://doi.org/10.1103/PhysRevB.105.195421} {\bibfield  {journal} {\bibinfo
   {journal} {Physical Review B}\ }\textbf {\bibinfo {volume} {105}},\ \bibinfo
  {pages} {195421} (\bibinfo {year} {2022})}\BibitemShut {NoStop}%
\bibitem [{\citenamefont {Pezo}\ \emph {et~al.}(2023)\citenamefont {Pezo},
  \citenamefont {Ovalle},\ and\ \citenamefont {Manchon}}]{pezo2023orbital}%
  \BibitemOpen
  \bibfield  {author} {\bibinfo {author} {\bibfnamefont {A.}~\bibnamefont
  {Pezo}}, \bibinfo {author} {\bibfnamefont {D.~G.}\ \bibnamefont {Ovalle}},\
  and\ \bibinfo {author} {\bibfnamefont {A.}~\bibnamefont {Manchon}},\
  }\href@noop {} {\bibinfo {title} {Orbital {H}all physics in two-dimensional
  {D}irac materials}} (\bibinfo {year} {2023}),\ \Eprint
  {https://arxiv.org/abs/2301.01126} {arXiv:2301.01126 [cond-mat.mes-hall]}
  \BibitemShut {NoStop}%
\bibitem [{\citenamefont {Yang}\ and\ \citenamefont
  {Nagaosa}(2014)}]{yang2014emergent}%
  \BibitemOpen
  \bibfield  {author} {\bibinfo {author} {\bibfnamefont {B.-J.}\ \bibnamefont
  {Yang}}\ and\ \bibinfo {author} {\bibfnamefont {N.}~\bibnamefont {Nagaosa}},\
  }\bibfield  {title} {\bibinfo {title} {Emergent topological phenomena in thin
  films of pyrochlore iridates},\ }\href
  {https://doi.org/10.1103/PhysRevLett.112.246402} {\bibfield  {journal}
  {\bibinfo  {journal} {Physical Review Letters}\ }\textbf {\bibinfo {volume}
  {112}},\ \bibinfo {pages} {246402} (\bibinfo {year} {2014})}\BibitemShut
  {NoStop}%
\bibitem [{Note1()}]{Note1}%
  \BibitemOpen
  \bibinfo {note} {We replaced `Im' (determining the imaginary part of the
  following expression) by the negative of the imaginary unit `$-\protect
  \mathrm {i}$'.}\BibitemShut {Stop}%
\bibitem [{\citenamefont {Go}\ \emph {et~al.}(2017)\citenamefont {Go},
  \citenamefont {Hanke}, \citenamefont {Buhl}, \citenamefont {Freimuth},
  \citenamefont {Bihlmayer}, \citenamefont {Lee}, \citenamefont {Mokrousov},\
  and\ \citenamefont {Bl{\"u}gel}}]{go2017toward}%
  \BibitemOpen
  \bibfield  {author} {\bibinfo {author} {\bibfnamefont {D.}~\bibnamefont
  {Go}}, \bibinfo {author} {\bibfnamefont {J.-P.}\ \bibnamefont {Hanke}},
  \bibinfo {author} {\bibfnamefont {P.~M.}\ \bibnamefont {Buhl}}, \bibinfo
  {author} {\bibfnamefont {F.}~\bibnamefont {Freimuth}}, \bibinfo {author}
  {\bibfnamefont {G.}~\bibnamefont {Bihlmayer}}, \bibinfo {author}
  {\bibfnamefont {H.-W.}\ \bibnamefont {Lee}}, \bibinfo {author} {\bibfnamefont
  {Y.}~\bibnamefont {Mokrousov}},\ and\ \bibinfo {author} {\bibfnamefont
  {S.}~\bibnamefont {Bl{\"u}gel}},\ }\bibfield  {title} {\bibinfo {title}
  {Toward surface orbitronics: giant orbital magnetism from the orbital
  {R}ashba effect at the surface of $sp$-metals},\ }\href
  {https://doi.org/10.1038/srep46742} {\bibfield  {journal} {\bibinfo
  {journal} {Scientific Reports}\ }\textbf {\bibinfo {volume} {7}},\ \bibinfo
  {pages} {46742} (\bibinfo {year} {2017})}\BibitemShut {NoStop}%
\bibitem [{\citenamefont {Yin}\ \emph {et~al.}(2018)\citenamefont {Yin},
  \citenamefont {Zhang}, \citenamefont {Li}, \citenamefont {Jiang},
  \citenamefont {Chang}, \citenamefont {Zhang}, \citenamefont {Lian},
  \citenamefont {Xiang}, \citenamefont {Belopolski}, \citenamefont {Zheng},
  \citenamefont {Cochran}, \citenamefont {Xu}, \citenamefont {Bian},
  \citenamefont {Liu}, \citenamefont {Chang}, \citenamefont {Lin},
  \citenamefont {Lu}, \citenamefont {Wang}, \citenamefont {Jia}, \citenamefont
  {Wang},\ and\ \citenamefont {Hasan}}]{Yin2018giant}%
  \BibitemOpen
  \bibfield  {author} {\bibinfo {author} {\bibfnamefont {J.-X.}\ \bibnamefont
  {Yin}}, \bibinfo {author} {\bibfnamefont {S.~S.}\ \bibnamefont {Zhang}},
  \bibinfo {author} {\bibfnamefont {H.}~\bibnamefont {Li}}, \bibinfo {author}
  {\bibfnamefont {K.}~\bibnamefont {Jiang}}, \bibinfo {author} {\bibfnamefont
  {G.}~\bibnamefont {Chang}}, \bibinfo {author} {\bibfnamefont
  {B.}~\bibnamefont {Zhang}}, \bibinfo {author} {\bibfnamefont
  {B.}~\bibnamefont {Lian}}, \bibinfo {author} {\bibfnamefont {C.}~\bibnamefont
  {Xiang}}, \bibinfo {author} {\bibfnamefont {I.}~\bibnamefont {Belopolski}},
  \bibinfo {author} {\bibfnamefont {H.}~\bibnamefont {Zheng}}, \bibinfo
  {author} {\bibfnamefont {T.~A.}\ \bibnamefont {Cochran}}, \bibinfo {author}
  {\bibfnamefont {S.-Y.}\ \bibnamefont {Xu}}, \bibinfo {author} {\bibfnamefont
  {G.}~\bibnamefont {Bian}}, \bibinfo {author} {\bibfnamefont {K.}~\bibnamefont
  {Liu}}, \bibinfo {author} {\bibfnamefont {T.-R.}\ \bibnamefont {Chang}},
  \bibinfo {author} {\bibfnamefont {H.}~\bibnamefont {Lin}}, \bibinfo {author}
  {\bibfnamefont {Z.-Y.}\ \bibnamefont {Lu}}, \bibinfo {author} {\bibfnamefont
  {Z.}~\bibnamefont {Wang}}, \bibinfo {author} {\bibfnamefont {S.}~\bibnamefont
  {Jia}}, \bibinfo {author} {\bibfnamefont {W.}~\bibnamefont {Wang}},\ and\
  \bibinfo {author} {\bibfnamefont {M.~Z.}\ \bibnamefont {Hasan}},\ }\bibfield
  {title} {\bibinfo {title} {Giant and anisotropic many-body spin--orbit
  tunability in a strongly correlated kagome magnet},\ }\href
  {https://doi.org/10.1038/s41586-018-0502-7} {\bibfield  {journal} {\bibinfo
  {journal} {Nature}\ }\textbf {\bibinfo {volume} {562}},\ \bibinfo {pages}
  {91} (\bibinfo {year} {2018})}\BibitemShut {NoStop}%
\bibitem [{\citenamefont {Khadka}\ \emph {et~al.}(2020)\citenamefont {Khadka},
  \citenamefont {Thapaliya}, \citenamefont {Hurtado~Parra}, \citenamefont
  {Wen}, \citenamefont {Need}, \citenamefont {Kikkawa},\ and\ \citenamefont
  {Huang}}]{Khadka2020anomalous}%
  \BibitemOpen
  \bibfield  {author} {\bibinfo {author} {\bibfnamefont {D.}~\bibnamefont
  {Khadka}}, \bibinfo {author} {\bibfnamefont {T.~R.}\ \bibnamefont
  {Thapaliya}}, \bibinfo {author} {\bibfnamefont {S.}~\bibnamefont
  {Hurtado~Parra}}, \bibinfo {author} {\bibfnamefont {J.}~\bibnamefont {Wen}},
  \bibinfo {author} {\bibfnamefont {R.}~\bibnamefont {Need}}, \bibinfo {author}
  {\bibfnamefont {J.~M.}\ \bibnamefont {Kikkawa}},\ and\ \bibinfo {author}
  {\bibfnamefont {S.~X.}\ \bibnamefont {Huang}},\ }\bibfield  {title} {\bibinfo
  {title} {Anomalous {H}all and {N}ernst effects in epitaxial films of
  topological kagome magnet {Fe$_{3}$}{Sn$_{2}$}},\ }\href
  {https://doi.org/10.1103/PhysRevMaterials.4.084203} {\bibfield  {journal}
  {\bibinfo  {journal} {Physical Review Materials}\ }\textbf {\bibinfo {volume}
  {4}},\ \bibinfo {pages} {084203} (\bibinfo {year} {2020})}\BibitemShut
  {NoStop}%
\bibitem [{\citenamefont {Liu}\ \emph {et~al.}(2018)\citenamefont {Liu},
  \citenamefont {Sun}, \citenamefont {Kumar}, \citenamefont {Muechler},
  \citenamefont {Sun}, \citenamefont {Jiao}, \citenamefont {Yang},
  \citenamefont {Liu}, \citenamefont {Liang}, \citenamefont {Xu}, \citenamefont
  {Kroder}, \citenamefont {S{\"u}{\ss}}, \citenamefont {Borrmann},
  \citenamefont {Shekhar}, \citenamefont {Wang}, \citenamefont {Xi},
  \citenamefont {Wang}, \citenamefont {Schnelle}, \citenamefont {Wirth},
  \citenamefont {Chen}, \citenamefont {Goennenwein},\ and\ \citenamefont
  {Felser}}]{Liu2018giant}%
  \BibitemOpen
  \bibfield  {author} {\bibinfo {author} {\bibfnamefont {E.}~\bibnamefont
  {Liu}}, \bibinfo {author} {\bibfnamefont {Y.}~\bibnamefont {Sun}}, \bibinfo
  {author} {\bibfnamefont {N.}~\bibnamefont {Kumar}}, \bibinfo {author}
  {\bibfnamefont {L.}~\bibnamefont {Muechler}}, \bibinfo {author}
  {\bibfnamefont {A.}~\bibnamefont {Sun}}, \bibinfo {author} {\bibfnamefont
  {L.}~\bibnamefont {Jiao}}, \bibinfo {author} {\bibfnamefont {S.-Y.}\
  \bibnamefont {Yang}}, \bibinfo {author} {\bibfnamefont {D.}~\bibnamefont
  {Liu}}, \bibinfo {author} {\bibfnamefont {A.}~\bibnamefont {Liang}}, \bibinfo
  {author} {\bibfnamefont {Q.}~\bibnamefont {Xu}}, \bibinfo {author}
  {\bibfnamefont {J.}~\bibnamefont {Kroder}}, \bibinfo {author} {\bibfnamefont
  {V.}~\bibnamefont {S{\"u}{\ss}}}, \bibinfo {author} {\bibfnamefont
  {H.}~\bibnamefont {Borrmann}}, \bibinfo {author} {\bibfnamefont
  {C.}~\bibnamefont {Shekhar}}, \bibinfo {author} {\bibfnamefont
  {Z.}~\bibnamefont {Wang}}, \bibinfo {author} {\bibfnamefont {C.}~\bibnamefont
  {Xi}}, \bibinfo {author} {\bibfnamefont {W.}~\bibnamefont {Wang}}, \bibinfo
  {author} {\bibfnamefont {W.}~\bibnamefont {Schnelle}}, \bibinfo {author}
  {\bibfnamefont {S.}~\bibnamefont {Wirth}}, \bibinfo {author} {\bibfnamefont
  {Y.}~\bibnamefont {Chen}}, \bibinfo {author} {\bibfnamefont {S.~T.~B.}\
  \bibnamefont {Goennenwein}},\ and\ \bibinfo {author} {\bibfnamefont
  {C.}~\bibnamefont {Felser}},\ }\bibfield  {title} {\bibinfo {title} {Giant
  anomalous {H}all effect in a ferromagnetic kagome-lattice semimetal},\ }\href
  {https://doi.org/10.1038/s41567-018-0234-5} {\bibfield  {journal} {\bibinfo
  {journal} {Nature Physics}\ }\textbf {\bibinfo {volume} {14}},\ \bibinfo
  {pages} {1125} (\bibinfo {year} {2018})}\BibitemShut {NoStop}%
\bibitem [{\citenamefont {Yazyev}(2019)}]{Yazyev2019an}%
  \BibitemOpen
  \bibfield  {author} {\bibinfo {author} {\bibfnamefont {O.~V.}\ \bibnamefont
  {Yazyev}},\ }\bibfield  {title} {\bibinfo {title} {An upside-down magnet},\
  }\href {https://doi.org/10.1038/s41567-019-0451-6} {\bibfield  {journal}
  {\bibinfo  {journal} {Nature Physics}\ }\textbf {\bibinfo {volume} {15}},\
  \bibinfo {pages} {424} (\bibinfo {year} {2019})}\BibitemShut {NoStop}%
\bibitem [{\citenamefont {Chen}\ \emph {et~al.}(2014)\citenamefont {Chen},
  \citenamefont {Niu},\ and\ \citenamefont {MacDonald}}]{chen2014anomalous}%
  \BibitemOpen
  \bibfield  {author} {\bibinfo {author} {\bibfnamefont {H.}~\bibnamefont
  {Chen}}, \bibinfo {author} {\bibfnamefont {Q.}~\bibnamefont {Niu}},\ and\
  \bibinfo {author} {\bibfnamefont {A.~H.}\ \bibnamefont {MacDonald}},\
  }\bibfield  {title} {\bibinfo {title} {Anomalous {H}all effect arising from
  noncollinear antiferromagnetism},\ }\href
  {https://doi.org/10.1103/PhysRevLett.112.017205} {\bibfield  {journal}
  {\bibinfo  {journal} {Physical Review Letters}\ }\textbf {\bibinfo {volume}
  {112}},\ \bibinfo {pages} {017205} (\bibinfo {year} {2014})}\BibitemShut
  {NoStop}%
\bibitem [{\citenamefont {Nakatsuji}\ \emph {et~al.}(2015)\citenamefont
  {Nakatsuji}, \citenamefont {Kiyohara},\ and\ \citenamefont
  {Higo}}]{nakatsuji2015large}%
  \BibitemOpen
  \bibfield  {author} {\bibinfo {author} {\bibfnamefont {S.}~\bibnamefont
  {Nakatsuji}}, \bibinfo {author} {\bibfnamefont {N.}~\bibnamefont
  {Kiyohara}},\ and\ \bibinfo {author} {\bibfnamefont {T.}~\bibnamefont
  {Higo}},\ }\bibfield  {title} {\bibinfo {title} {Large anomalous {H}all
  effect in a non-collinear antiferromagnet at room temperature},\ }\href
  {https://doi.org/10.1038/nature15723} {\bibfield  {journal} {\bibinfo
  {journal} {Nature}\ }\textbf {\bibinfo {volume} {527}},\ \bibinfo {pages}
  {212} (\bibinfo {year} {2015})}\BibitemShut {NoStop}%
\bibitem [{\citenamefont {Nayak}\ \emph {et~al.}(2016)\citenamefont {Nayak},
  \citenamefont {Fischer}, \citenamefont {Sun}, \citenamefont {Yan},
  \citenamefont {Karel}, \citenamefont {Komarek}, \citenamefont {Shekhar},
  \citenamefont {Kumar}, \citenamefont {Schnelle}, \citenamefont {K{\"u}bler}
  \emph {et~al.}}]{nayak2016large}%
  \BibitemOpen
  \bibfield  {author} {\bibinfo {author} {\bibfnamefont {A.~K.}\ \bibnamefont
  {Nayak}}, \bibinfo {author} {\bibfnamefont {J.~E.}\ \bibnamefont {Fischer}},
  \bibinfo {author} {\bibfnamefont {Y.}~\bibnamefont {Sun}}, \bibinfo {author}
  {\bibfnamefont {B.}~\bibnamefont {Yan}}, \bibinfo {author} {\bibfnamefont
  {J.}~\bibnamefont {Karel}}, \bibinfo {author} {\bibfnamefont {A.~C.}\
  \bibnamefont {Komarek}}, \bibinfo {author} {\bibfnamefont {C.}~\bibnamefont
  {Shekhar}}, \bibinfo {author} {\bibfnamefont {N.}~\bibnamefont {Kumar}},
  \bibinfo {author} {\bibfnamefont {W.}~\bibnamefont {Schnelle}}, \bibinfo
  {author} {\bibfnamefont {J.}~\bibnamefont {K{\"u}bler}}, \emph {et~al.},\
  }\bibfield  {title} {\bibinfo {title} {Large anomalous {H}all effect driven
  by a nonvanishing {B}erry curvature in the noncolinear antiferromagnet
  {Mn$_3$Ge}},\ }\href {https://doi.org/10.1126/sciadv.1501870} {\bibfield
  {journal} {\bibinfo  {journal} {Science Advances}\ }\textbf {\bibinfo
  {volume} {2}},\ \bibinfo {pages} {e1501870} (\bibinfo {year}
  {2016})}\BibitemShut {NoStop}%
\bibitem [{\citenamefont {Zhang}\ \emph {et~al.}(2017)\citenamefont {Zhang},
  \citenamefont {Sun}, \citenamefont {Yang}, \citenamefont {{\v{Z}}elezn{\`y}},
  \citenamefont {Parkin}, \citenamefont {Felser},\ and\ \citenamefont
  {Yan}}]{zhang2017strong}%
  \BibitemOpen
  \bibfield  {author} {\bibinfo {author} {\bibfnamefont {Y.}~\bibnamefont
  {Zhang}}, \bibinfo {author} {\bibfnamefont {Y.}~\bibnamefont {Sun}}, \bibinfo
  {author} {\bibfnamefont {H.}~\bibnamefont {Yang}}, \bibinfo {author}
  {\bibfnamefont {J.}~\bibnamefont {{\v{Z}}elezn{\`y}}}, \bibinfo {author}
  {\bibfnamefont {S.~P.}\ \bibnamefont {Parkin}}, \bibinfo {author}
  {\bibfnamefont {C.}~\bibnamefont {Felser}},\ and\ \bibinfo {author}
  {\bibfnamefont {B.}~\bibnamefont {Yan}},\ }\bibfield  {title} {\bibinfo
  {title} {Strong anisotropic anomalous {H}all effect and spin {H}all effect in
  the chiral antiferromagnetic compounds {Mn$_3$X} ({X = Ge, Sn, Ga, Ir, Rh,
  and Pt})},\ }\href {https://doi.org/10.1103/PhysRevB.95.075128} {\bibfield
  {journal} {\bibinfo  {journal} {Physical Review B}\ }\textbf {\bibinfo
  {volume} {95}},\ \bibinfo {pages} {075128} (\bibinfo {year}
  {2017})}\BibitemShut {NoStop}%
\bibitem [{Sup()}]{SupplMat}%
  \BibitemOpen
  \href@noop {} {}\bibinfo {note} {See Supplemental Material at [url inserted
  by the publisher] for complementing figures.}\BibitemShut {Stop}%
\bibitem [{\citenamefont {Niimi}\ \emph {et~al.}(2005)\citenamefont {Niimi},
  \citenamefont {Matsui}, \citenamefont {Kambara}, \citenamefont {Tagami},
  \citenamefont {Tsukada},\ and\ \citenamefont {Fukuyama}}]{niimi2005scanning}%
  \BibitemOpen
  \bibfield  {author} {\bibinfo {author} {\bibfnamefont {Y.}~\bibnamefont
  {Niimi}}, \bibinfo {author} {\bibfnamefont {T.}~\bibnamefont {Matsui}},
  \bibinfo {author} {\bibfnamefont {H.}~\bibnamefont {Kambara}}, \bibinfo
  {author} {\bibfnamefont {K.}~\bibnamefont {Tagami}}, \bibinfo {author}
  {\bibfnamefont {M.}~\bibnamefont {Tsukada}},\ and\ \bibinfo {author}
  {\bibfnamefont {H.}~\bibnamefont {Fukuyama}},\ }\bibfield  {title} {\bibinfo
  {title} {Scanning tunneling microscopy and spectroscopy studies of graphite
  edges},\ }\href
  {https://doi.org/https://doi.org/10.1016/j.apsusc.2004.09.091} {\bibfield
  {journal} {\bibinfo  {journal} {Applied Surface Science}\ }\textbf {\bibinfo
  {volume} {241}},\ \bibinfo {pages} {43} (\bibinfo {year} {2005})},\ \bibinfo
  {note} {the 9th International Symposium on Advanced Physical
  Fields}\BibitemShut {NoStop}%
\bibitem [{\citenamefont {Kobayashi}\ \emph {et~al.}(2005)\citenamefont
  {Kobayashi}, \citenamefont {Fukui}, \citenamefont {Enoki}, \citenamefont
  {Kusakabe},\ and\ \citenamefont {Kaburagi}}]{Kobayashi2005observation}%
  \BibitemOpen
  \bibfield  {author} {\bibinfo {author} {\bibfnamefont {Y.}~\bibnamefont
  {Kobayashi}}, \bibinfo {author} {\bibfnamefont {K.-i.}\ \bibnamefont
  {Fukui}}, \bibinfo {author} {\bibfnamefont {T.}~\bibnamefont {Enoki}},
  \bibinfo {author} {\bibfnamefont {K.}~\bibnamefont {Kusakabe}},\ and\
  \bibinfo {author} {\bibfnamefont {Y.}~\bibnamefont {Kaburagi}},\ }\bibfield
  {title} {\bibinfo {title} {Observation of zigzag and armchair edges of
  graphite using scanning tunneling microscopy and spectroscopy},\ }\href
  {https://doi.org/10.1103/PhysRevB.71.193406} {\bibfield  {journal} {\bibinfo
  {journal} {Physical Review B}\ }\textbf {\bibinfo {volume} {71}},\ \bibinfo
  {pages} {193406} (\bibinfo {year} {2005})}\BibitemShut {NoStop}%
\bibitem [{\citenamefont {Sasaki}\ \emph {et~al.}(2006)\citenamefont {Sasaki},
  \citenamefont {Murakami},\ and\ \citenamefont
  {Saito}}]{Sasaki2006stabilization}%
  \BibitemOpen
  \bibfield  {author} {\bibinfo {author} {\bibfnamefont {K.}~\bibnamefont
  {Sasaki}}, \bibinfo {author} {\bibfnamefont {S.}~\bibnamefont {Murakami}},\
  and\ \bibinfo {author} {\bibfnamefont {R.}~\bibnamefont {Saito}},\ }\bibfield
   {title} {\bibinfo {title} {{Stabilization mechanism of edge states in
  graphene}},\ }\href {https://doi.org/10.1063/1.2181274} {\bibfield  {journal}
  {\bibinfo  {journal} {Applied Physics Letters}\ }\textbf {\bibinfo {volume}
  {88}},\ \bibinfo {pages} {113110} (\bibinfo {year} {2006})}\BibitemShut
  {NoStop}%
\bibitem [{\citenamefont {Yao}\ \emph {et~al.}(2009)\citenamefont {Yao},
  \citenamefont {Yang},\ and\ \citenamefont {Niu}}]{Yao2009Edgestates}%
  \BibitemOpen
  \bibfield  {author} {\bibinfo {author} {\bibfnamefont {W.}~\bibnamefont
  {Yao}}, \bibinfo {author} {\bibfnamefont {S.~A.}\ \bibnamefont {Yang}},\ and\
  \bibinfo {author} {\bibfnamefont {Q.}~\bibnamefont {Niu}},\ }\bibfield
  {title} {\bibinfo {title} {Edge states in graphene: From gapped flat-band to
  gapless chiral modes},\ }\href
  {https://doi.org/10.1103/PhysRevLett.102.096801} {\bibfield  {journal}
  {\bibinfo  {journal} {Physical Review Letters}\ }\textbf {\bibinfo {volume}
  {102}},\ \bibinfo {pages} {096801} (\bibinfo {year} {2009})}\BibitemShut
  {NoStop}%
\bibitem [{\citenamefont {Costa}\ \emph {et~al.}(2023)\citenamefont {Costa},
  \citenamefont {Focassio}, \citenamefont {Canonico}, \citenamefont {Cysne},
  \citenamefont {Schleder}, \citenamefont {Muniz}, \citenamefont {Fazzio},\
  and\ \citenamefont {Rappoport}}]{costa2023connecting}%
  \BibitemOpen
  \bibfield  {author} {\bibinfo {author} {\bibfnamefont {M.}~\bibnamefont
  {Costa}}, \bibinfo {author} {\bibfnamefont {B.}~\bibnamefont {Focassio}},
  \bibinfo {author} {\bibfnamefont {L.~M.}\ \bibnamefont {Canonico}}, \bibinfo
  {author} {\bibfnamefont {T.~P.}\ \bibnamefont {Cysne}}, \bibinfo {author}
  {\bibfnamefont {G.~R.}\ \bibnamefont {Schleder}}, \bibinfo {author}
  {\bibfnamefont {R.~B.}\ \bibnamefont {Muniz}}, \bibinfo {author}
  {\bibfnamefont {A.}~\bibnamefont {Fazzio}},\ and\ \bibinfo {author}
  {\bibfnamefont {T.~G.}\ \bibnamefont {Rappoport}},\ }\bibfield  {title}
  {\bibinfo {title} {Connecting higher-order topology with the orbital {H}all
  effect in monolayers of transition metal dichalcogenides},\ }\href
  {https://doi.org/10.1103/PhysRevLett.130.116204} {\bibfield  {journal}
  {\bibinfo  {journal} {Physical Review Letters}\ }\textbf {\bibinfo {volume}
  {130}},\ \bibinfo {pages} {116204} (\bibinfo {year} {2023})}\BibitemShut
  {NoStop}%
\bibitem [{\citenamefont {Sun}\ \emph {et~al.}(2022)\citenamefont {Sun},
  \citenamefont {Chen}, \citenamefont {Du}, \citenamefont {Chen}, \citenamefont
  {Zhou},\ and\ \citenamefont {Ye}}]{Sun2022spinvalley}%
  \BibitemOpen
  \bibfield  {author} {\bibinfo {author} {\bibfnamefont {Y.-L.}\ \bibnamefont
  {Sun}}, \bibinfo {author} {\bibfnamefont {G.-H.}\ \bibnamefont {Chen}},
  \bibinfo {author} {\bibfnamefont {S.-C.}\ \bibnamefont {Du}}, \bibinfo
  {author} {\bibfnamefont {Z.-B.}\ \bibnamefont {Chen}}, \bibinfo {author}
  {\bibfnamefont {Y.-W.}\ \bibnamefont {Zhou}},\ and\ \bibinfo {author}
  {\bibfnamefont {E.-J.}\ \bibnamefont {Ye}},\ }\bibfield  {title} {\bibinfo
  {title} {Spin-valley polarized edge states in quasi-one-dimensional
  asymmetric kagome lattice},\ }\href
  {https://doi.org/10.3389/fphy.2022.1033836} {\bibfield  {journal} {\bibinfo
  {journal} {Frontiers in Physics}\ }\textbf {\bibinfo {volume} {10}},\
  \bibinfo {pages} {33836} (\bibinfo {year} {2022})}\BibitemShut {NoStop}%
\bibitem [{\citenamefont {Jo}\ \emph {et~al.}(2018)\citenamefont {Jo},
  \citenamefont {Go},\ and\ \citenamefont {Lee}}]{jo2018gigantic}%
  \BibitemOpen
  \bibfield  {author} {\bibinfo {author} {\bibfnamefont {D.}~\bibnamefont
  {Jo}}, \bibinfo {author} {\bibfnamefont {D.}~\bibnamefont {Go}},\ and\
  \bibinfo {author} {\bibfnamefont {H.-W.}\ \bibnamefont {Lee}},\ }\bibfield
  {title} {\bibinfo {title} {Gigantic intrinsic orbital {H}all effects in
  weakly spin-orbit coupled metals},\ }\href
  {https://doi.org/10.1103/PhysRevB.98.214405} {\bibfield  {journal} {\bibinfo
  {journal} {Physical Review B}\ }\textbf {\bibinfo {volume} {98}},\ \bibinfo
  {pages} {214405} (\bibinfo {year} {2018})}\BibitemShut {NoStop}%
\bibitem [{\citenamefont {Choi}\ \emph {et~al.}(2023)\citenamefont {Choi},
  \citenamefont {Jo}, \citenamefont {Ko}, \citenamefont {Go}, \citenamefont
  {Kim}, \citenamefont {Park}, \citenamefont {Kim}, \citenamefont {Min},
  \citenamefont {Choi},\ and\ \citenamefont {Lee}}]{Choi2023observation}%
  \BibitemOpen
  \bibfield  {author} {\bibinfo {author} {\bibfnamefont {Y.-G.}\ \bibnamefont
  {Choi}}, \bibinfo {author} {\bibfnamefont {D.}~\bibnamefont {Jo}}, \bibinfo
  {author} {\bibfnamefont {K.-H.}\ \bibnamefont {Ko}}, \bibinfo {author}
  {\bibfnamefont {D.}~\bibnamefont {Go}}, \bibinfo {author} {\bibfnamefont
  {K.-H.}\ \bibnamefont {Kim}}, \bibinfo {author} {\bibfnamefont {H.~G.}\
  \bibnamefont {Park}}, \bibinfo {author} {\bibfnamefont {C.}~\bibnamefont
  {Kim}}, \bibinfo {author} {\bibfnamefont {B.-C.}\ \bibnamefont {Min}},
  \bibinfo {author} {\bibfnamefont {G.-M.}\ \bibnamefont {Choi}},\ and\
  \bibinfo {author} {\bibfnamefont {H.-W.}\ \bibnamefont {Lee}},\ }\bibfield
  {title} {\bibinfo {title} {Observation of the orbital {H}all effect in a
  light metal {T}i},\ }\href {https://doi.org/10.1038/s41586-023-06101-9}
  {\bibfield  {journal} {\bibinfo  {journal} {Nature}\ }\textbf {\bibinfo
  {volume} {619}},\ \bibinfo {pages} {52} (\bibinfo {year} {2023})}\BibitemShut
  {NoStop}%
\bibitem [{\citenamefont {Seifert}\ \emph {et~al.}(2023)\citenamefont
  {Seifert}, \citenamefont {Go}, \citenamefont {Hayashi}, \citenamefont
  {Rouzegar}, \citenamefont {Freimuth}, \citenamefont {Ando}, \citenamefont
  {Mokrousov},\ and\ \citenamefont {Kampfrath}}]{seifert2023time}%
  \BibitemOpen
  \bibfield  {author} {\bibinfo {author} {\bibfnamefont {T.~S.}\ \bibnamefont
  {Seifert}}, \bibinfo {author} {\bibfnamefont {D.}~\bibnamefont {Go}},
  \bibinfo {author} {\bibfnamefont {H.}~\bibnamefont {Hayashi}}, \bibinfo
  {author} {\bibfnamefont {R.}~\bibnamefont {Rouzegar}}, \bibinfo {author}
  {\bibfnamefont {F.}~\bibnamefont {Freimuth}}, \bibinfo {author}
  {\bibfnamefont {K.}~\bibnamefont {Ando}}, \bibinfo {author} {\bibfnamefont
  {Y.}~\bibnamefont {Mokrousov}},\ and\ \bibinfo {author} {\bibfnamefont
  {T.}~\bibnamefont {Kampfrath}},\ }\bibfield  {title} {\bibinfo {title}
  {Time-domain observation of ballistic orbital-angular-momentum currents with
  giant relaxation length in tungsten},\ }\bibfield  {journal} {\bibinfo
  {journal} {Nature Nanotechnology}\ }\href
  {https://doi.org/10.1038/s41565-023-01470-8} {10.1038/s41565-023-01470-8}
  (\bibinfo {year} {2023})\BibitemShut {NoStop}%
\bibitem [{\citenamefont {Bhowal}\ and\ \citenamefont
  {Vignale}(2021)}]{bhowal2021orbital}%
  \BibitemOpen
  \bibfield  {author} {\bibinfo {author} {\bibfnamefont {S.}~\bibnamefont
  {Bhowal}}\ and\ \bibinfo {author} {\bibfnamefont {G.}~\bibnamefont
  {Vignale}},\ }\bibfield  {title} {\bibinfo {title} {Orbital {H}all effect as
  an alternative to valley {H}all effect in gapped graphene},\ }\href
  {https://doi.org/10.1103/PhysRevB.103.195309} {\bibfield  {journal} {\bibinfo
   {journal} {Physical Review B}\ }\textbf {\bibinfo {volume} {103}},\ \bibinfo
  {pages} {195309} (\bibinfo {year} {2021})}\BibitemShut {NoStop}%
\bibitem [{\citenamefont {Cysne}\ \emph {et~al.}(2021)\citenamefont {Cysne},
  \citenamefont {Costa}, \citenamefont {Canonico}, \citenamefont {Nardelli},
  \citenamefont {Muniz},\ and\ \citenamefont
  {Rappoport}}]{cysne2021disentangling}%
  \BibitemOpen
  \bibfield  {author} {\bibinfo {author} {\bibfnamefont {T.~P.}\ \bibnamefont
  {Cysne}}, \bibinfo {author} {\bibfnamefont {M.}~\bibnamefont {Costa}},
  \bibinfo {author} {\bibfnamefont {L.~M.}\ \bibnamefont {Canonico}}, \bibinfo
  {author} {\bibfnamefont {M.~B.}\ \bibnamefont {Nardelli}}, \bibinfo {author}
  {\bibfnamefont {R.}~\bibnamefont {Muniz}},\ and\ \bibinfo {author}
  {\bibfnamefont {T.~G.}\ \bibnamefont {Rappoport}},\ }\bibfield  {title}
  {\bibinfo {title} {Disentangling orbital and valley {H}all effects in
  bilayers of transition metal dichalcogenides},\ }\href
  {https://doi.org/10.1103/PhysRevLett.126.056601} {\bibfield  {journal}
  {\bibinfo  {journal} {Physical Review Letters}\ }\textbf {\bibinfo {volume}
  {126}},\ \bibinfo {pages} {056601} (\bibinfo {year} {2021})}\BibitemShut
  {NoStop}%
\end{thebibliography}%

\end{document}